\documentclass[journal,twoside,twocolumn]{IEEEtran}
\pdfoutput=1  

\usepackage{amsmath,amssymb}
\usepackage{graphicx}
\usepackage{xcolor}
\usepackage{cite}
\usepackage{balance}
\usepackage{algorithmic}
\usepackage{algorithm}

\usepackage{arydshln} 

\usepackage{tikz}
\usepackage{pgfplots}
\pgfplotsset{compat=1.14} 
\usetikzlibrary{arrows,positioning,shadows,shapes,intersections}

\usepackage{hyperref}
\hypersetup{
  breaklinks=true,  
  colorlinks=true,  
  citecolor=blue,  
  linkcolor=blue  
}

\renewcommand\arraystretch{1.3}  
\newcommand\tstrut{\rule{0pt}{2.5ex}} 

\newcommand{\R}{{\mathbb{R}}}
\newcommand{\Z}{{\mathbb{Z}}}
\newcommand{\T}{{\mathrm{T}}}

\newcommand{\ba}{{\boldsymbol{a}}}

\newcommand{\bA}{{\boldsymbol{A}}}

\newcommand{\bB}{{\boldsymbol{B}}}
\newcommand{\tB}{\boldsymbol{\tilde{B}}}

\newcommand{\bg}{{\boldsymbol{g}}}

\newcommand{\bI}{{\boldsymbol{I}}}
\newcommand{\bM}{{\boldsymbol{M}}}
\newcommand{\bn}{{\boldsymbol{n}}}

\newcommand{\bu}{{\boldsymbol{u}}}
\newcommand{\bU}{{\boldsymbol{U}}}
\newcommand{\bbU}{\bar{\bU}}
\newcommand{\bv}{{\boldsymbol{v}}}

\newcommand{\bW}{{\boldsymbol{W}}}
\newcommand{\bx}{{\boldsymbol{x}}}

\newcommand{\bzero}{{\boldsymbol{0}}}

\newcommand{\polytope}{\Omega'}
\newcommand{\grp}{{\mathcal{G}}}
\newcommand{\verts}{{\mathcal{V}}}
\newcommand{\normals}{\mathcal{N}}
\newcommand{\facet}{\mathcal{F}}

\newcommand{\h}{\tfrac{1}{2}} 
\newcommand{\qh}{\tfrac{q}{2}} 
\newcommand{\ah}{\tfrac{a}{2}} 
\newcommand{\dx}{\,\mathrm{d}\bx}
\newcommand{\Iphasea}{P_{a}}
\newcommand{\Iphase}{P_{\ba}}

\newcommand{\IphaseEpsA}{P_{\epsilon,\ba}}
\newcommand{\defeq}{\triangleq}

\newcommand{\LL}{L} 
\newcommand{\gL}{\LL'} 

\newcommand{\Lp}{\LL_\textrm{p}} 

\newcommand{\sL}{\LL} 
\newcommand{\uL}{\bar\LL} 
\newcommand{\guL}{\bar\LL'} 
\newcommand{\sgL}{\LL'} 

\newcommand{\sg}{\bg} 
\newcommand{\ug}{\bar\bg} 
\newcommand{\sGamma}{\Gamma} 
\newcommand{\uGamma}{\bar\Gamma} 

\newcommand{\baopt}{{\boldsymbol{a}_{\text{opt}}}}
\newcommand{\baoptA}{\ba_B}
\newcommand{\aopt}{a_{\text{opt}}}

\newcommand{\AEnine}{\text{AE}_9}

\DeclareMathOperator{\tr}{tr} 
\DeclareMathOperator{\Aut}{Aut}
\DeclareMathOperator{\Diag}{diag}

\title{Parametric Lattices Are Better Quantizers \\ in Dimensions 13 and 14}
\author{%
    Daniel Pook-Kolb,
    Erik Agrell, \IEEEmembership{Fellow, IEEE},
    and
    Bruce Allen, \IEEEmembership{Member, IEEE}
    \thanks{The work of E.~Agrell was supported by a Collaborating Scientist
        Grant from the Max Planck Institute for Gravitational Physics, Germany,
        which is gratefully acknowledged.}%
    \thanks{D.~Pook-Kolb and B.~Allen are with the
        Max Planck Institute for Gravitational Physics, 30167 Hannover, Germany,
        and
        Leibniz Universit\"at Hannover
        (e-mail: daniel.pook.kolb@aei.mpg.de and bruce.allen@aei.mpg.de).}
    \thanks{E.~Agrell is with the
        Department of Electrical Engineering, Chalmers University of Technology,
        41296 Gothenburg, Sweden
        (e-mail: agrell@chalmers.se).}
}

\begin{document}

\maketitle

\begin{abstract} 
    New lattice quantizers with lower
    normalized second moments than previously reported
    are constructed in 13 and 14 dimensions and conjectured to be optimal.
    Our construction combines an initial numerical optimization
    with a subsequent
    analytical optimization of
    families of lattices,
    whose Voronoi regions are constructed exactly.
    The new lattices are constructed from
    glued products of previously known lattices, by
    scaling the component lattices and
    then optimizing the scale factors.
    A two-parameter family of lattices in 13 dimensions reveals an intricate
    landscape of phase changes as the parameters are varied.
\end{abstract}


\section{Introduction}
\label{sec:intro}

\IEEEPARstart{T}{he}
\emph{lattice quantizer} problem is to find a lattice with the minimal
normalized second moment in a given dimension \cite[Ch.~2]{conway99splag}.
Geometrically, this minimizes the mean squared distance from a random point
to the closest lattice point, when the number of lattice points per unit
volume is fixed.
It is one of the classical problems in lattice theory and has applications in
 fields ranging from
data compression \cite{gray98}, \cite[Chs.~1, 3--5]{zamir14book}
and cryptography \cite{guo15}
to machine learning \cite{sablayrolles19}
and data analysis \cite{allen21prd}.

Proofs of optimality of a lattice for quantization exist only in dimensions up
to three.
In two dimensions, the optimal lattice quantizer is the hexagonal lattice
\cite{fejestoth59} and in three dimensions
it is the body-centered cubic lattice \cite{barnes83}.
Together with these proven results,
the lattices with the lowest known normalized second moments
in dimensions $4$--$8$, $12$, $16$, and $24$
were listed in 1984 by Conway and Sloane \cite{conway84} and
conjectured optimal by the same authors in 1999~\cite[p.~13]{conway99splag}.
Remarkably, the conjectured optimal lattice quantizer in $12$ dimensions, the
Coxeter--Todd lattice $K_{12}$, is surpassed by lattices found in 2010 by
Kudryashov and Yurkov \cite{Kudryashov10} and rediscovered and analyzed
    by the present authors 
    using a different construction technique \cite{agrell24K12};
    see Appendix~\ref{app:equal12}.

In dimensions $13$ and $14$, the focus of the present paper, the best known
lattice quantizers were for a long time the duals of the checkerboard lattice
$D^*_{13}$ and $D^*_{14}$ \cite[Fig.~2]{conway84}, \cite[Fig.~2.9]{conway99splag}.
Both were surpassed by a product of $K_{12}$ and a low-dimensional lattice \cite{agrell23}.
Using a better quantizer than $K_{12}$ in $12$ dimensions gives
a slightly improved product lattice in $13$ dimensions \cite{agrell24K12}.
A lamination of $K_{12}$ led to a non-product lattice with an even lower NSM \cite{Pook-Kolb2022Exact}.
Lyu et al.\ obtained a better $14$-dimensional lattice from a complex lattice
over the Eisenstein integers in 2023 \cite{lyu22}.

With one exception, the best known lattice quantizers are \emph{integral lattices,}
i.e., lattices such that the inner product between any pair of lattice vectors is an integer.
The exception is in $9$ dimensions, where a conjectured optimal lattice
was found in 1998 by Agrell and Eriksson using numerical optimization \cite{agrell98}.
It was defined using a $9\times 9$ generator matrix, where $80$ components are
exact and one is ``just a computer number,'' as the lattice pioneer Neil Sloane once
commented \cite{sloane98personal}. It took until 2021 to identify the square of this ``computer number''
as the root of a ninth-order polynomial \cite{allen21}.
We call such lattices, which are exact functions
of a small number of real parameters, \emph{parametric lattices.}

The work reported here employs both numerical approximation methods and
exact analytic ones. While this mixed approach to finding optimal lattice quantizers is decades old, there has been recent progress.
In \cite{agrell24opt}, numerical optimization was performed in dimensions
$10$--$16$ using an improved
stochastic gradient descent method.
This is complemented in the same work by a new identification technique to
find exact lattices from approximate numerical representations.
In all dimensions except $13$, $14$, and $15$, the previously known lattices
with the lowest normalized second moment
were recovered, providing strong numerical evidence for the conjecture that these are
the optimal lattice quantizers in their dimension.
In dimension $15$, the lattice $\Lambda^*_{15}$, i.e., the dual of the
laminated lattice $\Lambda_{15}$, was identified and conjectured optimal.
Results in the only remaining dimensions $13$ and $14$ were reported
numerically, with further analysis deferred to future work.
That is the purpose of the present paper.

First, the lattices found numerically in dimensions $13$ and $14$ are
identified using the technique described in \cite{agrell24opt}:
solving systems of equations derived from their theta images.
The technique does not return exact representations of these two lattices,
but rather reveals them to be parametric lattices,
somewhat similar to the best $9$-dimensional one \cite{allen21}.
Then, by optimizing the parameters using the methods developed in
\cite{allen21, Pook-Kolb2022Exact}, we obtain exact lattice quantizers with the lowest
normalized second moments reported in dimensions $13$ and $14$,
and conjecture these lattices to be optimal in their dimensions.

\section{Lattice Definitions}

\subsection{Basic notions}
\label{sec:basics}

An $n$-dimensional lattice $\LL$ in $\R^m$, $m \ge n$, is the set of all linear
combinations of $n$ linearly independent {\em basis vectors} in $\R^m$ with integer
coefficients.
With the convention that all vectors are row vectors, we can write
\begin{equation}\label{eq:lattice}
    \LL \defeq \left\{ \bu \bB \colon \bu \in \Z^n \right\},
\end{equation}
where $\bB$ is an $n \times m$ {\em generator matrix} of $\LL$ whose rows
are the basis vectors.
There are infinitely many generator matrices for the same lattice $\LL$.

The {\em dual lattice} $\LL^*$ of a lattice $\LL$ with generator matrix $\bB$
is generated by the matrix $\bB^* \defeq \bA^{-1}\bB$, where
$\bA \defeq \bB\bB^T$ is a {\em Gram matrix} of $\LL$.
Similar to generator matrices, Gram matrices are not unique for a given
lattice, but their determinants depend only on $\LL$ \cite[p.~4]{conway99splag}.
A lattice is an integral lattice if and only if its Gram matrix has only
integer entries.

Two lattices $\LL$ and $\LL'$ are called {\em equivalent} if one can be
transformed into the other by a combination of rotations, reflections, and/or changes of scale
\cite[p.~10]{conway99splag}.
This means that for any $n$-dimensional lattice $\LL$ in $\R^m$, $m > n$,
there exists an equivalent lattice $\LL'$ in $\R^n$.
To simplify the discussion and without loss of generality, we henceforth assume
that $m = n$, making $\bB$ a square generator matrix.%
\footnote{This notational assumption will however not prevent us from using
nonsquare submatrices as building blocks in the square generator matrices
\eqref{eq:gen13plain} and \eqref{eq:gen13}.}
In that case, $\bB^* = (\bB^T)^{-1}$.

The {\em Voronoi region} of a lattice point $\bx \in \LL$ is the set of all
points in $\R^n$ closer to $\bx$ than to any other lattice point.
By definition, the origin $\bzero$ is always a lattice point, and we denote
the Voronoi region of $\bzero$ by $\Omega$.
The volume of $\Omega$ is $V = \sqrt{\det\bA} = \lvert\det\bB\rvert$.

The lattice quantizer problem is to find a lattice
$\LL$ with the minimal {\em normalized second moment} (NSM) or
{\em quantizer constant}
\cite[pp.~34, 56--62]{conway99splag}
\begin{equation}\label{eq:NSM}
    G(\LL) \defeq \frac{1}{nV^{1+2/n}} \int_\Omega \|\bx\|^2 \dx.
\end{equation}
The normalization by $1/V^{1+2/n}$ makes $G$ independent of the
lattice's scale, i.e., $G(\LL) = G(a\LL)$ for all $a>0$.
The factor of $1/n$ is a convention that ensures that the $n$-dimensional
cubic lattice $\Z^n$ has the same NSM in all dimensions, with $G(\Z^n) = 1/12$.
It also has the consequence that
\begin{equation}\label{eq:G-of-product}
    G(\LL \times \LL) = G(\LL),
\end{equation}
where $\LL \times \LL$ is a product lattice defined in
Sec.~\ref{sec:product}.
If $\LL$ and $\LL'$ are equivalent lattices, then their respective NSMs are
the same, $G(\LL) = G(\LL')$.

In 1996, Zamir and Feder \cite{zamir96} proved a useful necessary condition
that a globally optimal lattice quantizer satisfies.
Let $\bU$ denote the {\em second moment matrix} of $\LL$ defined by
\begin{equation}\label{eq:Uab}
    \bU \defeq \frac{1}{V} \int_\Omega \bx^T \bx \dx.
\end{equation}
Then, if $\LL$ is a globally optimal lattice quantizer,
$\bU$ is proportional to the $n$-dimensional identity matrix $\bI_n$.
This result is generalized in \cite{agrell23}
to apply to {\em locally} optimal lattice quantizers as well.
Note that this is only a necessary condition.
If $\bU$ is not proportional to the identity, then 
$\LL$
is not even locally optimal.
This will become an important point in our analysis, especially for the novel
$13$-dimensional lattice.

\subsection{Product lattices}
\label{sec:product}

The Cartesian product of $k$ scaled lattices $\LL_1,\ldots,\LL_k$ is
\begin{equation}\label{eq:scaledprod}
    \Lp \defeq a_1 \LL_1 \times \cdots \times a_k \LL_k
        = \{ [a_1 \bx_1\;\cdots\;a_k \bx_k] \colon \bx_i \in \LL_i, \forall i \}
    ,
\end{equation}
where $a_1, \ldots, a_k$ are real positive scale factors.
$\Lp$ is called a {\em product lattice} and $\LL_1, \ldots, \LL_k$ the
(possibly scaled) {\em component lattices}.
It is a special case of a parametric lattice with $k$ parameters $a_1, \ldots, a_k$.
The dimension of $\Lp$ is $n = n_1 +\cdots+ n_k$, where $n_i$ is the dimension of $\LL_i$.

Product lattices were first considered for quantization in \cite{gersho79}.
It was shown in \cite{agrell23}
that there exist optimal choices of factors $a_i$ in \eqref{eq:scaledprod} that
result in the lowest possible NSM of the product lattice.
Such an {\em optimal product lattice} has an NSM satisfying
\begin{equation}\label{eq:prod-opt-NSM}
    G(\Lp)^n = G(\LL_1)^{n_1} \cdots G(\LL_k)^{n_k}
    .
\end{equation}
By combining various lattices in this way and selecting the result with
the lowest NSM for a given dimension $n$, many new lattices were found with
lower NSMs than previously reported \cite{agrell23,lyu22}.

A crucial result in \cite{agrell23} is that a product lattice~\eqref{eq:scaledprod}
is never optimal.
There always exists a small perturbation of
the generator matrix of a product lattice that reduces the NSM,
so that product lattices are not even locally optimal.

\subsection{Glued lattices}
\label{sec:gluing}

The concept of ``gluing'' a lattice was originally developed by
Conway and Sloane in \cite{conway82jnt}
and later considered for quantization in \cite{agrell24K12}.
It becomes especially useful when applied to product lattices, where it has
led to two different $12$-dimensional lattice quantizers with a lower NSM than the
Coxeter--Todd lattice $K_{12}$.

Intuitively, gluing interpolates between a lattice $\LL$ and its dual $\LL^*$,
\begin{equation}\label{eq:interp}
    \LL \subseteq \gL \subseteq \LL^*,
\end{equation}
where $\gL$ is the glued lattice.
Since this requires that $\LL \subseteq \LL^*$, it implies that $\LL$ must be
an integral lattice.
In general, the dual $\LL^*$ of an integral lattice
will not be an integral lattice unless $\LL$ is self-dual.

Since an integral lattice $\LL$ is contained in its dual $\LL^*$,
the dual can be written as a finite number of unions of its translates
\begin{equation}\label{eq:L*-as-union}
    \LL^* = \bigcup_{\bg\in\LL^*/\LL} (\LL + \bg).
\end{equation}
The quotient $\LL^*/\LL$ is a largest set of vectors in $\LL^*$
that are all distinct modulo translations by
vectors in $\LL$.
This set is not unique, but possible choices differ only by translations in $\LL$
and therefore leave \eqref{eq:L*-as-union} invariant.
Conventionally, the vectors are selected as short as possible within their
coset $\LL + \bg$, and by this convention, $\bzero$ is always one of the vectors.
Thus,
$\LL^*/\LL$ is a finite set of coset representatives of $\LL$ in
$\LL^*$ \cite[p.~48]{conway99splag}.
The representatives $\bg$ in \eqref{eq:L*-as-union} form a group
under addition modulo $\LL$ and
are called the {\em glue vectors} of $\LL$
The number of glue vectors, i.e., the index of $\LL$ in $\LL^*$, is
$\det\bA$.

We use the term ``gluing'' when only a subset of glue vectors is used in
\eqref{eq:L*-as-union},
\begin{equation}\label{eq:gluing}
    \gL \defeq \bigcup_{\bg\in\Gamma} (\LL + \bg),
\end{equation}
where $\Gamma \subseteq \LL^*/\LL$ is a group under addition
modulo $\LL$, called the {\em glue group}.
(If $\Gamma$ is not a group, then \eqref{eq:gluing} produces a nonlattice
packing.)

A product lattice $\Lp$ with integral component lattices $\LL_i$,
$i = 1,\ldots,k$, and $a_i = 1$ for all $i$, is itself an integral lattice and can thus be used for
gluing.
The glue vectors of $\Lp$ consist of concatenations of the glue vectors of
$\LL_i$,
\begin{equation}\label{eq:glue-words}
    \bg = [\bg_1 \; \ldots \; \bg_k],
\end{equation}
where $\bg_i \in \LL_i^* / \LL_i$, and are called {\em glue words}.

\subsection{Gluing of scaled lattices}
\label{sec:scaledgluing}

We generalize the gluing construction \eqref{eq:gluing} by applying it to
product lattices of the form \eqref{eq:scaledprod} for arbitrary positive
$a_i$, where the (unscaled) component lattices are integral
lattices.
For this, a useful identity that holds for any lattice $\LL$ and scale factor
$a > 0$ is that
\begin{equation}\label{eq:aLDual}
    (a\LL)^* = a^{-1} \LL^*,
\end{equation}
which follows from the definition of the dual.

First, we define the gluing of scaled integral lattices.
Let $\uL$ be an integral lattice
and
$\sL = a\uL$ for some $a>0$.
In general, $\sL \not\subseteq \sL^*$, so it cannot be glued as defined in Sec.~\ref{sec:gluing}
since $\sL^* / \sL$ does not exist.
However,
by \eqref{eq:aLDual} we have $\sL^* = (a\uL)^* = a^{-1}\uL^*$
and thus $\sL = a\uL \subseteq a\uL^* = a^2\sL^*$.
This means that there exist cosets $(a^2\sL^*) / \sL$ and
\begin{equation}\label{eq:scaled-L*}
    a^2 \sL^* = \bigcup_{\sg\in (a^2\sL^*)/\sL} (\sL + \sg).
\end{equation}
The glue vectors $\sg$ are the coset representatives in
\begin{equation}\label{eq:general-glue-vectors}
    (a^2\sL^*)/\sL = (a\uL^*)/(a\uL) = a \, \left( \uL^*/\uL \right)
\end{equation}
(the factor $a$ does not cancel since ``$/$'' forms the quotient space)
and are therefore simply the usual glue vectors of $\uL$
multiplied by the scale factor $a$.
If $\uGamma$ is a glue group of $\uL$ and $\guL$ the corresponding
glued lattice obtained from \eqref{eq:gluing}, then we define the
glued {\em scaled} lattice via
\begin{equation}\label{eq:scaled-gluing}
    \sgL \defeq \bigcup_{\sg\in \sGamma} (\sL + \sg)
    = \bigcup_{\ug\in \uGamma} (a\uL + a\ug)
    = a \guL,
\end{equation}
where the glue group $\sGamma=a\uGamma$.

For a product lattice $\Lp$ of the form \eqref{eq:scaledprod}, i.e.,
\begin{equation}\label{eq:scaledprod-bar}
    \Lp = a_1 \uL_1 \times \ldots \times a_k \uL_k,
\end{equation}
with $k$
integral component lattices $\uL_1, \ldots, \uL_k$,
we define gluing by composing glue words using the
glue vectors of the scaled component lattices,
\begin{equation}\label{eq:glued-scaled-product}
    \gL \defeq
        \bigcup_{\bg \in \Gamma} (\Lp + \bg) =
        \bigcup_{[\ug_1\;\ldots\;\ug_k] \in \uGamma} (\Lp + [a_1\ug_1 \; \ldots \; a_k\ug_k])
    .
\end{equation}
In \eqref{eq:glued-scaled-product}, the glue group $\Gamma$
is a subset of
\begin{equation}\label{eq:scaled-product-glue-group}
    a_1 \, (\uL_1^*/\uL_1) \times \ldots \times a_k \, (\uL_k^*/\uL_k)
\end{equation}
and is a group under addition modulo $\Lp$.
The group $\uGamma$ is a subset of
\begin{equation}\label{eq:general-glue-group}
    \uL_1^*/\uL_1 \times \ldots \times \uL_k^*/\uL_k
\end{equation}
and a group under addition modulo the unscaled lattice
$\uL_1 \times \cdots \times \uL_k$.
Note that $\gL$ is not a sublattice of the dual $\Lp^*$ or of a
scaled version of $\Lp^*$, unless the individual factors $a_i$ are all equal.

\subsection{Generalized gluing}
\label{sec:gengluing}

The gluing constructions \eqref{eq:gluing} and \eqref{eq:glued-scaled-product}
can be generalized further as follows. We no longer require any
relationship between $\Gamma$ and glue vectors of any constituent lattice.
If $\Gamma$ is a finite group under addition modulo $\LL$, then
\eqref{eq:gluing} still produces a lattice $\gL$ containing $\LL$.
This general case was already mentioned in \cite[Sec.~III]{agrell24K12} but not
considered further.
It contains the gluing of products of scaled integral lattices as
a special case but also permits glue vectors not in $\LL^*$ or non-integral $\LL$.

\section{Parametric Voronoi Regions}
\label{sec:voronoi}

We analyze lattices using the methods
described in \cite{Pook-Kolb2022Exact} with the optimizations in \cite{pook-kolb23}.
This is done by exactly constructing the full Voronoi region of the lattice up
to symmetry.
Using the recursion relations of \cite[Sec.~3]{Pook-Kolb2022Exact},
which build upon \cite[Sec.~3]{allen21},
the NSM \eqref{eq:NSM} and second moment matrix \eqref{eq:Uab}
can then be calculated.

The Voronoi region of an $n$-dimensional lattice $\LL$ is represented as a {\em hierarchy} of faces and
subfaces, starting with the single $n$-dimensional face, the Voronoi region
$\Omega$, at the top with no parents
(see, e.g., \cite[Fig.~4]{Pook-Kolb2022Exact}).
The facets are the $(n-1)$-dimensional subfaces of $\Omega$.
Their only parent face is $\Omega$.
These facets are, like $\Omega$, convex polytopes and each have $(n-2)$-dimensional
subfaces (i.e., their respective facets), which are the $(n-2)$-faces of $\Omega$.
Our method builds the complete face hierarchy of $\Omega$ from
the Voronoi region itself down to the $0$-dimensional faces, which are the vertices of $\Omega$.
This hierarchy is sometimes called the {\em Hasse diagram.}

The automorphism group $\Aut(\LL)$ of the lattice is a subgroup of the
rotations $O(n)$ that takes $\LL$ into $\LL$.
If $\bB$ is a generator of $\LL$ and $\bM \in \Aut(\LL)$,
then $\bB\bM$ is another generator of the same lattice $\LL$.
Symmetries are exploited in the following way.
Since $\Aut(\LL)$ consists of the symmetries of $\Omega$, it takes each
$d$-dimensional face of $\Omega$ into another $d$-dimensional face of
$\Omega$.
Two faces of $\Omega$ are called {\em equivalent} under $\Aut(\LL)$ if an
element of $\Aut(\LL)$ takes one into the other.
The hierarchy of faces we construct now only includes one face per class of
equivalent faces, and additionally all their direct subfaces.
(The latter are required to identify all classes of faces and to evaluate the
recursion relations.)

Some issues are unique to analyzing a parametric lattice.
Let
$a_1, \ldots, a_s$ be its $s$ real parameters, $s \geq 1$, and $\ba \defeq [a_1, \ldots, a_s]$.
The generator matrix of a parametric lattice is then a function of the
parameters, and we write it as $\bB(\ba)$.
Let further $\Omega(\ba)$ be the Voronoi region of the lattice generated by $\bB(\ba)$.
As the parameters approach certain critical points,
vertices of $\Omega(\ba)$ may split into multiple vertices, or
several vertices may merge into one or disappear.
Such topological changes have to be considered when constructing a
representation of the Voronoi region.

Our method builds the face hierarchy at a fixed point $\ba$ in parameter space.
We obtain the positions of all vertices,
relevant vectors, and other quantities like the volumes of all faces and the NSM
as functions of the parameters.
However, since the face hierarchy is assumed fixed, these functions are valid
only when $\Omega(\ba)$ undergoes no topological changes.
If $\Omega(\ba)$ does undergo a topological change, our
representation does not agree anymore with $\Omega(\ba)$ and may, in fact, become
geometrically ill-defined.

To account for topological changes of the Voronoi region, we
determine the region $\Iphase \subset \R^s$ that contains $\ba$ and where
the Voronoi region has the same topology throughout.
The region $\Iphase$ is called the {\em phase} of $\ba$.
An example where phases of a parametric lattice have been analyzed can be found in
\cite[Sec.~4]{allen21}.
In particular, \cite[Tab.~3]{allen21} shows how the numbers of vertices and
facets of the Voronoi region of the parametric lattice $\AEnine$ change across phases.

The phase $\Iphase$ is identified in two steps \cite{allen21,Pook-Kolb2022Exact}.
First, we identify the region $\Iphase'$, with $\ba \in \Iphase'$, where the lengths of all
$1$-faces are positive
(recall that those lengths,
i.e., the volumes of the $1$-faces, are functions of $\ba$).
If the length of a $1$-face vanishes, then two vertices merge and
the face hierarchy must change.
The second step is to test $\Omega$ for further topological
changes in $\Iphase'$ as described in the next paragraph.
If no change occurs, then $\Iphase' = \Iphase$.
Otherwise, the boundaries of $\Iphase$ are located and the phase is determined as
the interior region
$\Iphase \subset \Iphase'$.
For the lattices studied here, no such changes were found away from
points of vanishing $1$-face lengths,
and thus for these lattices $\Iphase' = \Iphase$.

We use the following tests to identify topological changes of $\Omega$ within
$\Iphase'$.
Each test is carried out at a large number of points
(a few hundred for the lattices studied here)
$\ba' \in \Iphase'$.\footnote{%
This works only if the number $s$ of parameters is small.
In cases of many parameters, testing the conditions on a dense enough
set of points in $\Iphase'$ might be too expensive.}
\begin{itemize}
    \item[($i$)]
        The set of parametric facets is the same throughout $\Iphase'$.
        This is tested by writing the relevant vectors $\normals(\ba)$ of $\Omega(\ba)$ as
        $\bu\bB(\ba)$, for integer vectors $\bu$, and checking that
        $\bu\bB(\ba')$ are the relevant vectors of $\Omega(\ba')$ for the same
        set of integer vectors $\bu$.
        This requires finding all relevant vectors once at each test point $\ba'$
        and writing them as linear combinations of the rows of
        $\bB(\ba')$.
    \item[($ii$)]
        Parametric vertices do not move outside of $\Omega$.
        More precisely,
        let $\verts_{\ba}$ be the set of parametric vertices of $\Omega(\ba)$ and
        \begin{equation}\label{eq:vertsparam}
            \verts_{\ba}(\ba') \defeq \left\{ \bv(\ba') \colon \bv(\ba) \in \verts_{\ba} \right\}
            .
        \end{equation}
        Then we test if $\verts_{\ba}(\ba') \subset \Omega(\ba')$
        as follows.
        The set of parametric vertices $\verts_{\ba}$ is constructed once at $\ba$.
        Each vertex is then evaluated at $\ba'$, which produces the set $\verts_{\ba}(\ba')$.
        With the relevant vectors $\normals(\ba')$ known (see ($i$)), the set of
        inequalities that defines $\Omega(\ba')$ is evaluated, i.e., we test
        \begin{equation}\label{eq:OmegaIneq}
            \bv \cdot \bn \le \frac{1}{2} \|\bn\|^2,
            \ \ \forall \bv \in \verts_{\ba}(\ba'), \bn \in \normals(\ba').
        \end{equation}
    \item[($iii$)]
        Vertices do not move outside the facets they lie in; i.e.,
        if a vertex $\bv(\ba)$ lies in a facet $\facet(\ba)$, then $\bv(\ba') \in \facet(\ba')$.
        This is tested as in ($ii$), but instead of checking for inequality in
        \eqref{eq:OmegaIneq}, it is verified that equality holds at $\ba'$
        if and only if it holds at $\ba$.
        In practice, ($ii$) and ($iii$) are checked together.
    \item[($iv$)]
        The volume of the convex hull of $\verts_{\ba}(\ba')$
        is equal to $V(\ba') = \lvert \det \bB(\ba') \rvert$.
        The volume of this convex hull is a byproduct of
        our calculations, making the test very inexpensive.
\end{itemize}
If condition ($i$) is not satisfied at $\ba'$, then the set of faces has changed
and our face hierarchy does not represent $\Omega(\ba')$.
Conditions ($ii$) and ($iii$) together ensure that
$\verts_{\ba}(\ba')$ are vertices of $\Omega(\ba')$.
The face hierarchy is then well-defined and represents a convex polytope $\polytope$
contained in $\Omega(\ba')$.
Condition ($iv$) ensures that its volume agrees with the volume $V(\ba')$ of $\Omega(\ba')$,
which then implies $\polytope = \Omega(\ba')$.

\section{Lattice Optimization}
\label{sec:optimization}

An algorithm to numerically optimize lattices was presented in \cite{agrell24opt}.
It works by iteratively
updating the generator matrix towards lower NSMs,
guided by stochastic gradient descent. Thus, the algorithm has similarities with
standard machine learning techniques, although without any black-box elements.
It works without explicitly determining the Voronoi region, which is an
extremely time- and memory-consuming operation in high dimensions.

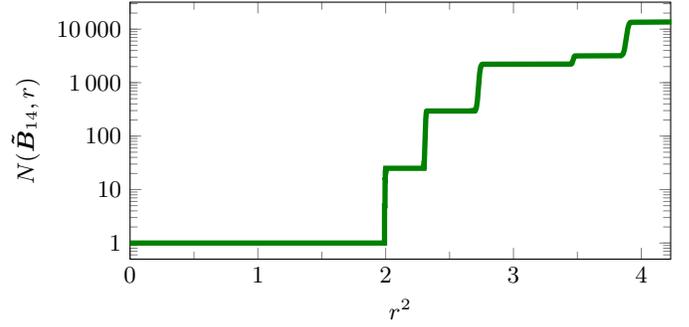
\begin{figure}\centering
    \begin{tikzpicture}
        \sffamily\small
        \begin{axis}[
            width=.99\columnwidth, height=5cm,
            xmin=0, xmax=4.23,
            minor x tick num=1,
            xlabel={$r^2$},
            xlabel near ticks,
            ymode=log,
            log ticks with fixed point,
            ymin=0.5, ymax=32000,
            ylabel={$N(\tB_{14},r)$},
            ylabel near ticks,
            ytick distance=10,
            yticklabel style={/pgf/number format/1000 sep=\,}, 
            legend style={at={(.96,.08)}, anchor=south east, legend cell align=left, font=\footnotesize}
        ]
        \addplot[green!50!black, line width=2] table{fig-data/logth14.txt};
        \end{axis}
    \end{tikzpicture}
    \caption{%
        Theta image of a numerically optimized $14$-dimensional lattice.
    }
    \label{f:theta14}
\end{figure}

The output of the numerical optimization algorithm is an approximation of a locally
optimal lattice. Therefore,
a heuristic method to identify the exact form of a lattice whose
generator or Gram matrix is known only numerically was also proposed in
\cite{agrell24opt}. It exploits the regular structure that almost all locally
optimal lattices exhibit, specifically the arrangement of lattice points on a discrete
set of separated, spherical shells.

To characterize, visualize, and identify these shells,
the cumulative distribution of lattice point norms
\begin{equation}\label{eq:theta}
    N(\bB, r) \defeq \left|\{ \bu \in \Z^n \colon \|\bu \bB\| \leq r \}\right|
\end{equation}
is used, where $\bB$ is the generator matrix of the lattice.
The plot of $N(\bB, r)$ as a function of $r^2$ is called the {\em theta image}
of $\bB$ and an example for the $14$-dimensional numerical result
$\tB_{14}$ (i.e., with approximate entries)
 of \cite[Sec.~VI]{agrell24opt} is shown in Fig.~\ref{f:theta14}.
Vertical steps in this plot, which represent the shells of lattice points with almost
equal norms, are identified visually. They are turned into an
overdetermined system of equations, which is then solved analytically using
common mathematical software.
This procedure is described in detail in \cite[Sec.~V-B]{agrell24opt}.
Three possible outcomes can occur now:
the system has one, multiple, or no solution.
If it has one solution, then the exact lattice is recovered.
The case of multiple solutions corresponds to a parametric lattice.
In the rare cases when the target lattice
does not have a structure of well separated shells, the identification technique fails.
The first
such case we have observed appears in $13$ dimensions and will be
described in detail in Sec.~\ref{sec:13D}.

In the remainder of this section, we introduce a deterministic algorithm
to fine-tune and exactly identify a locally optimal lattice from a close
approximation, without relying on any presumed lattice structure.
It is very demanding in terms of computer time and memory, and we
apply it only as a last resort when the identification method
in \cite{agrell24opt} fails to return an exact, locally optimal lattice.

A second moment matrix $\bU$ that is not proportional to the identity can be used to
find a deformation $\bA_\epsilon$ of any lattice $\LL$ with generator $\bB$
towards a lower NSM.
This idea is introduced in the proof of \cite[Theorem~2]{agrell23}.
Let
\begin{equation}\label{eq:Utraceless}
    \bbU \defeq \bU - \frac{\tr\bU}{n} \bI_n
\end{equation}
be the traceless part of $\bU$,
which is nonzero if $\bU$ is not proportional to $\bI_n$,
and $\bA_\epsilon \defeq \exp(-\epsilon\bbU)$.
It is then shown in \cite{agrell23}
that for small enough positive $\epsilon$, the lattice $\LL\bA_\epsilon$
with generator $\bB \bA_\epsilon$ has a lower NSM than $\LL$.

An optimization algorithm based on this insight operates as follows.
Starting from an arbitrary lattice $\LL$ with generator matrix $\bB$,
the full face hierarchy of the Voronoi region, from dimension $0$ to $n$, is computed.
From this description, the NSM $G$ and the second moment matrix $\bU$
are computed exactly.
A perturbation matrix $\bA_\epsilon = \exp(-\epsilon\bbU)$ is computed with $\bbU$
given by \eqref{eq:Utraceless}.
An alternative is $\bA_\epsilon = \bI_n-\epsilon\bbU$,
which is approximately equivalent to $\exp(-\epsilon\bbU)$ 
if the elements of $\epsilon\bbU$ have small enough magnitude.
Then the generator matrix $\bB$ is perturbed into $\bB \bA_\epsilon$.
This yields a parametric lattice $\LL \bA_\epsilon$, with
$\epsilon$ as a parameter.

To select $\epsilon$, we consider the
deformed Voronoi region $\Omega\bA_\epsilon$, where $\Omega$ is the Voronoi
region of $\LL$. This region is different from the Voronoi region of $\LL\bA_\epsilon$,
but it is easy to compute from $\Omega$ when $\bA_\epsilon$ is known.
The second moment matrix of $\Omega\bA_\epsilon$ is $\bA_\epsilon^T \bU \bA_\epsilon$
\cite[Theorem~2]{agrell23}. In general, there exists no perturbation $\epsilon$ that makes
$\bA_\epsilon^T \bU \bA_\epsilon$ proportional to $\bI_n$,
but $\epsilon$ can nevertheless be chosen to move
the off-diagonal elements closer to zero and the diagonal elements closer to each other.
This provides
a good enough approximation for the next step.

At this point it may be useful to introduce additional parameters into the
parametric lattice.
This is not strictly necessary but can significantly reduce the number of
steps needed for the optimization algorithm to converge.
It is motivated by the fact that there may be no value of $\epsilon$ for which
$\LL\bA_\epsilon$ has a second moment matrix proportional to $\bI_n$.
An example is the $13$-dimensional case studied in Sec.~\ref{sec:13D}, which
is a generalized glued product of three integral lattices.
There, a natural choice of additional parameters are the scale factors
$\ba = [a_1,a_2,a_3]$ of the component lattices.

Next, the face hierarchy of $\LL \bA_{\epsilon}$ is computed for the selected
$\epsilon$.
This yields a complete description of the
Voronoi region at $\epsilon$ and within the phase $\IphaseEpsA$.
This is why the exact value of $\epsilon$ is not critical.
The NSM is computed as a function of $\epsilon$ and any additional parameters in the phase
and the minimum is found analytically. If this minimum lies in $\IphaseEpsA$,
then the algorithm has converged; otherwise the search continues in another phase.
There are at least three methods to select the next phase to investigate.

One is to consider the phase in which the minimum of the NSM expression of $\IphaseEpsA$ lies.
Although this expression does not give the NSM in other phases,
it is often a very accurate approximation. For example, \eqref{eq:NSM13B}
shows that the second moment of the $13$-dimensional parametric lattice in
Sec.~\ref{sec:13D} is $13$ times differentiable on the boundary between the
two analyzed phases.

The second method is to compute the second moment matrix of $\LL \bA_{\epsilon}$
and again perturb the lattice to move this matrix closer to a scaled identity.
With this and the previous method, it needs to be verified that the exact NSM at the new point is
less than the previous NSM.

If this is not the case (which is theoretically possible although we have
not experienced it in practice), then a third, safer, method is to
numerically find the minimum NSM in $\IphaseEpsA$, which
by assumption lies on the phase boundary, and next investigate the other
phases that touch the boundary at the same point, proceeding
with the ones containing the smallest NSM.

We successfully used the first of these three options to optimize the
$13$-dimensional lattice studied here.
This required computing the face hierarchy at the initial point and in
two additional phases.
The $14$-dimensional lattice was optimized in just a single step, since its
second moment matrix was proportional to $\bI_n$ within the initial phase.

\section{Dimension 14: Glued $K'_{10} \times D_4$}
\label{sec:14D}

The new lattices studied in this paper were initially found numerically in
\cite{agrell24opt}.
In dimensions $10$--$16$, the gradient descent method numerically minimized the
NSM, starting from random lattices.
After $100$ such searches were performed in each dimension,
the lattice with the lowest NSM was selected for further analysis.
In all of these dimensions except $13$ and $14$, the system of equations obtained from
the theta image has a unique solution and the resulting lattices can thus be
represented by an exact generator matrix $\bB$.

This section analyzes the previously unidentified lattice in dimension $14$.
We start with the $14$-dimensional lattice as its analysis is less
complicated and establishes the basic steps upon which the analysis of the
$13$-dimensional lattice then builds.

\subsection{Identification of the lattice}
\label{sec:14Dident}

The system of equations built from the theta image of the numerically found $14$-dimensional lattice
in Fig.~\ref{f:theta14}
is underdetermined, regardless of how many shells are considered.
Although the Gram matrix cannot be uniquely
determined at this stage, it contains only one unknown, real parameter.
Thus, the new $14$-dimensional lattice is a parametric lattice with
one degree of freedom.

A generator matrix was obtained from the Gram matrix by Cholesky decomposition,
and the generator matrix
was further refined via basis changes, rotations, and scaling, aiming to confine the
influence of the real parameter cleanly to as few dimensions as possible.
The parametric generator matrix
\begin{equation}\label{eq:gen14}
    \begin{aligned}[b]
    \bB_{14}(a) = \left[
        \begin{array}{@{\!\!}*{10}{@{\;}c@{\;}};{2pt/2pt}@{}c@{}*{4}{@{\;}c@{\;}}@{\!\!}}
            2  & 0    & 0  & 0    & 0  & 0    & 0  & 0   & 0  & 0   && 0   & 0   & 0   & 0 \\
            1  & q    & 0  & 0    & 0  & 0    & 0  & 0   & 0  & 0   && 0   & 0   & 0   & 0 \\
            0  & 0    & 2  & 0    & 0  & 0    & 0  & 0   & 0  & 0   && 0   & 0   & 0   & 0 \\
            0  & 0    & 1  & q    & 0  & 0    & 0  & 0   & 0  & 0   && 0   & 0   & 0   & 0 \\
            0  & 0    & 0  & 0    & 2  & 0    & 0  & 0   & 0  & 0   && 0   & 0   & 0   & 0 \\
            0  & 0    & 0  & 0    & 1  & q    & 0  & 0   & 0  & 0   && 0   & 0   & 0   & 0 \\
            1  & 0    & \h & -\qh & \h & -\qh & 1  & 0   & 0  & 0   && 0   & 0   & 0   & 0 \\
            \h & \qh  & 1  & 0    & 1  & 0    & \h & \qh & 0  & 0   && 0   & 0   & 0   & 0 \\
            \h & -\qh & 1  & 0    & \h & -\qh & 0  & 0   & 1  & 0   && 0   & 0   & 0   & 0 \\
            \rule[-1.5ex]{0pt}{0pt}
            1  & 0    & \h & \qh  & 1  & 0    & 0  & 0   & \h & \qh && 0   & 0   & 0   & 0 \\
            \hdashline[2pt/2pt]
            \tstrut
            1  & 0    & 1  & 0    & \h & \qh  & 0  & 0   & 0  & 0   && a   & 0   & 0   & 0 \\
            0  & 0    & 0  & 0    & 0  & 0    & 0  & 0   & 0  & 0   && -a  & a   & 0   & 0 \\
            0  & 0    & 0  & 0    & 0  & 0    & 0  & 0   & 0  & 0   && -a  & 0   & a   & 0 \\
            \h & -\qh & \h & -\qh & 1  & 0    & 0  & 0   & 0  & 0   && \ah & \ah & \ah & \ah \\
        \end{array}
    \right]
    \\[-.5ex]
    \end{aligned}
\end{equation}
was obtained, where $q \defeq \sqrt{3}$ and $a$ is the unknown parameter.

The lattice generated by \eqref{eq:gen14} may be obtained by
gluing scaled integral lattices.
We set $k=2$, $\uL_1=K'_{10}$, $\uL_2=D_4$, $a_1=1$, and $a_2=a$
in \eqref{eq:scaledprod-bar} and \eqref{eq:glued-scaled-product} to obtain
\begin{equation}\label{eq:prod14}
    \gL \defeq
        \bigcup_{[\bg_1\ \bg_2] \in \Gamma} (K'_{10}+\bg_1) \times (a D_4 + \bg_2)
        .
\end{equation}
Here, $K'_{10}$
is a laminated $10$-dimensional lattice
\cite[Sec.~8.5]{martinet03}, which is equivalent to the lattice with automorphism group
$(C_6\times \mathrm{SU}_4(2))\colon C_2$ described in \cite{souvignier94},%
\footnote{\label{fn:10d}%
Their equivalence is proved in Appendix \ref{app:equal10}.}
and $D_4$ is the $4$-dimensional checkerboard lattice.
We can for example take the top-left $10\times 10$ submatrix of \eqref{eq:gen14}
as a generator matrix for $K'_{10}$
and \cite[Eq.~(36)]{agrell98} as a generator matrix for $D_4$.
With these choices, the glue group $\Gamma$ may be represented by the four vectors
\begin{equation}\label{eq:glue14}
        \begin{array}{c*{15}{@{\:\:}c}}
\big[ & 0 & 0 & 0 & 0 & 0 & 0 & 0 & 0 & 0 & 0 & 0 & 0 & 0 & 0 & \big],\\
\big[ & 1 & 0 & 1 & 0 & \h & \qh & 0 & 0 & 0 & 0 & a & 0 & 0 & 0 & \big],\\
\big[ & \h & -\qh & \h & -\qh & 1 & 0 & 0 & 0 & 0 & 0 & \ah & \ah & \ah & \ah & \big],\\
\big[ & \h & \qh & \h & \qh & -\h & \qh & 0 & 0 & 0 & 0 & \ah & -\ah & -\ah & -\ah & \big].
        \end{array}
\end{equation}
The first $10$ components of each vector in \eqref{eq:glue14} form a subgroup of $\uL_1^*/\uL_1$,
and the last $4$ components of the same vectors form the whole group $(a^2\uL_2^*)/\uL_2$.
For $a=1$, the $14$-dimensional vectors \eqref{eq:glue14} therefore form a subgroup of
$\Lp^*/\Lp$ for the product lattice $\Lp = \uL_1\times\uL_2$.
Together, \eqref{eq:prod14} and \eqref{eq:glue14} define the same lattice
as \eqref{eq:gen14}.

\subsection{Construction of the Voronoi region}

Even though the parameter $a$ cannot be determined exactly from the system of equations
obtained from the theta image, it can still be identified approximately.
The theta image of $\bB_{14}(a)$, properly scaled, best matches Fig.~\ref{f:theta14} for
$a \approx 25/19$.

The Voronoi region of the lattice \eqref{eq:gen14} was constructed at
$a = 25/19 \approx 1.3158$ as described in Sec.~\ref{sec:voronoi}.
The phase $\Iphasea$ is given by
\begin{equation}\label{eq:phase14}
    \sqrt{5/3} < a < \sqrt{13/7}
    ,
\end{equation}
i.e., $1.291 \lesssim a \lesssim 1.363$.
Conditions ($i$)--($iv$) of Sec.~\ref{sec:voronoi} are fulfilled in this interval, but not outside.
At both of these boundaries, at least one class of $1$-faces has zero length:
the two vertices at their endpoints merge to one.

The automorphism group $\Aut(\LL)$ of the lattice \eqref{eq:gen14} has order
$59\,719\,680$.
It is a subgroup of the direct product $O(10) \times O(4)$, i.e.,
all symmetries $\bM \in \Aut(\LL)$ are of the form
\begin{equation}\label{eq:sym14form}
    \bM = \left[
        \begin{array}{@{\,}c*{1}{@{\;\;}c}@{\,}}
            \bM_{10 \times 10} & \bzero \\
            \bzero & \bM_{4 \times 4}
        \end{array}
    \right],
\end{equation}
where $\bM_{d \times d}$ is an orthogonal $d \times d$ matrix.
Let $\pi_{10} \colon \Aut(\LL) \to O(10)$ and $\pi_{4} \colon \Aut(\LL) \to O(4)$ be the
projections that map a symmetry $\bM \in \Aut(\LL)$ of the form \eqref{eq:sym14form}
to their corresponding $10 \times 10$ and $4 \times 4$ blocks
$\bM_{10\times10}$ and $\bM_{4\times4}$, respectively.
Then, $\pi_{10}(\Aut(\LL))$ is a group of order
$311\,040$ that is the symmetry group of $K'_{10}$, whose order
has been reported in \cite[lattice~\emph{KAPPA10$'$}]{nebe:catalog}.
Similarly, $\pi_{4}(\Aut(\LL))$ is a group of order $1\,152$ that is
the automorphism group of $D_4$ \cite[Ch.~4]{conway99splag}.
However, while the product
\begin{equation}\label{eq:prod_of_projections}
    \pi_{10}(\Aut(\LL)) \times \pi_{4}(\Aut(\LL))
\end{equation}
of these two groups
(which has order $358\,318\,080$)
is the automorphism group of the (unglued) product lattice $K'_{10} \times a D_4$,
it is {\em not} the automorphism group of the glued lattice \eqref{eq:gen14}.

The subgroup
\begin{equation}\label{eq:G10}
    \grp_{10} \defeq \pi_{4}^{-1}(\bI_4) = \left\{ \bM \in \Aut(\LL) \colon \pi_{4}(\bM) = \bI_4 \right\},
\end{equation}
which affects only the first $10$ components,
has order $51\,840$.
Therefore, $\grp_{10}$ is isomorphic to a {\em strict} subgroup of $\pi_{10}(\Aut(\LL))$.
For the following discussion, it is helpful to define the embeddings
\begin{equation}\label{eq:iso10_4}
    \begin{aligned}[b]
        i_{10}(\bM_{10 \times 10}) &\defeq \left[
            \begin{array}{@{\,}c*{1}{@{\;\;}c}@{\,}}\bM_{10 \times 10} & \bzero \\ \bzero & \bI_4\end{array}
        \right], \\
        i_{4}(\bM_{4 \times 4}) &\defeq \left[
            \begin{array}{@{\,}c*{1}{@{\;\;}c}@{\,}}\bI_{10} & \bzero \\ \bzero & \bM_{4 \times 4}\end{array}
        \right].
    \end{aligned}
\end{equation}
The group $\grp_{10}$ is then generated by the following transformations:
\begin{itemize}
    \item sign-changes of component pairs $(x_{2i-1}, x_{2i})$,
        where $x_j$ is the $j$th component of a lattice vector $\bx$ and
        $i = 1, \ldots, 5$,
    \item joint swapping of pairs
        $(x_1, x_2)$ with $(x_3, x_4)$
        and
        $(x_7, x_8)$ with $(x_9, x_{10})$,
    \item joint swapping of pairs
        $(x_3, x_4)$ with $(x_7, x_8)$
        and
        $(x_5, x_6)$ with $(x_9, x_{10})$,
    \item the image of the $10 \times 10$ matrix
        \begin{equation}\label{eq:Mrefl14}
            \left[
                \begin{array}{*{5}c}
                    \bI_2/2    & \bW    & \bW    & -\bI_2/2   & \bzero \\
                    \bW^T  & \bI_2/2    & -\bI_2/2   & \bW^T  & \bzero \\
                    \bW^T  & -\bI_2/2   & \bI_2/2    & \bW^T  & \bzero \\
                    -\bI_2/2   & \bW    & \bW    & \bI_2/2    & \bzero \\
                    \bzero & \bzero & \bzero & \bzero & \bI_2 \\
                \end{array}
            \right]
        \end{equation}
        under $i_{10}$,
        where
        \begin{equation}\label{eq:VW}
                \bW \defeq \frac{1}{4}\left[
                        \begin{array}{*{2}c}
                            1 & -\sqrt{3} \\
                            \sqrt{3} & 1
                        \end{array}
                    \right].
        \end{equation}
\end{itemize}

The symmetries of the automorphism group of $K'_{10}$ that are
missing from $\grp_{10}$
are generated by
\begin{equation}\label{eq:missing10_1}
    \bM_{10} = \left[
            \begin{array}{*{5}c}
                \bM_2  & \bzero  & \bzero  & \bzero & \bzero \\
                \bzero & \bzero  & \bM_2'' & \bzero & \bzero \\
                \bzero & \bM_2'' & \bzero  & \bzero & \bzero \\
                \bzero & \bzero  & \bzero  & \bM_2  & \bzero \\
                \bzero & \bzero  & \bzero  & \bzero & \bM_2' \\
            \end{array}
        \right]
\end{equation}
and
\begin{equation}\label{eq:missing10_2}
    \bM_{10}' = \left[
            \begin{array}{*{5}c}
                \bM_2  & \bzero & \bzero & \bzero & \bzero \\
                \bzero & \bzero & \bzero & \bzero & \bM_2 \\
                \bzero & \bM_2  & \bzero & \bzero & \bzero \\
                \bzero & \bzero & \bM_2  & \bzero & \bzero \\
                \bzero & \bzero & \bzero & \bM_2  & \bzero \\
            \end{array}
        \right],
\end{equation}
where
\begin{equation}\label{eq:SQdef}
    \begin{gathered}[b]
        \bM_2 \defeq \frac{1}{2}\left[
                \begin{array}{cc}
                    1         & -\sqrt{3} \\
                    -\sqrt{3} & -1
                \end{array}
            \right],
        \qquad
        \bM_2' \defeq \frac{1}{2}\left[
                \begin{array}{cc}
                    1        & \sqrt{3} \\
                    \sqrt{3} & -1
                \end{array}
            \right],\\[1ex]
        \bM_2'' \defeq \Diag(1, -1).
    \end{gathered}
\end{equation}
Note that
$i_{10}(\bM_{10})$
and
$i_{10}(\bM_{10}')$
are {\em not} symmetries of the glued lattice \eqref{eq:gen14}.

We can also describe the subgroup
$\grp_{4} \defeq \pi_{10}^{-1}(\bI_{10}) \subset \Aut(\LL)$ of symmetries
affecting only the last $4$ components.
It consists of all permutations and an even number of sign changes of the last
four components.
Together, these form a group of order $192$ that is isomorphic to a subgroup
of the automorphism group of $D_4$.
The remaining symmetries of $D_4$ are, however,
not found in $\grp_{4}$.
They consist of
the Hadamard matrix
\begin{equation}\label{eq:H4}
    \bM_4 \defeq \frac{1}{2} \left[
        \begin{array}{@{\,}c*{3}{@{\;\;}r}@{\,}}
            1 & 1 & 1 & 1 \\
            1 &-1 & 1 &-1 \\
            1 & 1 &-1 &-1 \\
            1 &-1 &-1 & 1 \\
        \end{array}
    \right]
\end{equation}
and
a sign change of the last component,
\begin{equation}\label{eq:sign-last}
    \bM_4' \defeq \Diag(1, 1, 1, -1).
\end{equation}

The subgroups $\grp_{10}$ and $\grp_{4}$ together form a group of order
$9\,953\,280$.
By combining the generators of $\grp_{10}$ with those of $\grp_{4}$ and the two matrices
\begin{equation}\label{eq:combined_syms}
    \left[
        \begin{array}{cc}
            \bM_{10} & \bzero \\
            \bzero & \bM_4
        \end{array}
    \right]
    \quad\text{and}\quad
    \left[
        \begin{array}{cc}
            \bM_{10}' & \bzero \\
            \bzero & \bM_4'
        \end{array}
    \right],
\end{equation}
we finally obtain the full group $\Aut(\LL)$.

In practice, we found this group by first forming the direct product of the known
automorphism groups of $K'_{10}$ and $D_4$.
As noted above, $\Aut(\LL)$ is a strict subgroup of this larger group.
We then used GAP \cite{GAPSoftware} to initially determine $\grp_{10}$ and $\grp_{4}$
and subsequently analyzed combinations of the symmetries of $K'_{10}$ and
$D_4$ missing from $\grp_{10}$ and $\grp_{4}$.

\subsection{Optimization of the NSM}

\begin{figure}\centering
    \includegraphics[width=0.9\linewidth]{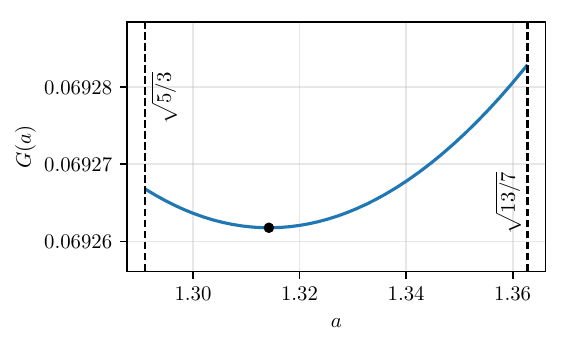}
    \caption{%
        NSM for the $14$-dimensional lattice \eqref{eq:gen14} as function of $a$.
        The dot marks the location of the minimum and the dashed vertical
        lines indicate the boundaries of the phase \eqref{eq:phase14}.
        The previously reported minimum NSM of $0.069\,52$
        \cite{lyu22} is about eight plot-heights above the dot.
    }
    \label{f:NSM14}
\end{figure}

The NSM as a function of $a$, $G(a)$, is given by \eqref{eq:NSM} with
$n=14$, $V(a) = \lvert\det\bB_{14}(a)\rvert = 9 \sqrt{3} a^4$, and
\begin{gather}\label{eq:U14}
    \begin{aligned}[b]
        \int_\Omega \|\bx\|^2\dx ={}&
            \frac{\sqrt{3}}{2^{24} \cdot 3^{13} \cdot 5^{6} \cdot 7^{4} \cdot 11^{2} \cdot 13 a^{4}} \\
            &\big(-203620624743138487 a^{30}\\
            &\phantom{\big(}+ 3801139170184845180 a^{28}\\
            &\phantom{\big(}- 12389539784482907235 a^{26}\\
            &\phantom{\big(}- 360690099386364296640 a^{24}\\
            &\phantom{\big(}+ 6016931591463402618645 a^{22}\\
            &\phantom{\big(}- 50328772679983887733644 a^{20}\\
            &\phantom{\big(}+ 281188164016693016348265 a^{18}\\
            &\phantom{\big(}- 1139254272224209154665080 a^{16}\\
            &\phantom{\big(}+ 3445025447618975371649355 a^{14}\\
            &\phantom{\big(}- 7776389251376810821296540 a^{12}\\
            &\phantom{\big(}+ 15825068372524956384278439 a^{10}\\
            &\phantom{\big(}+ 6097536730919383009631280 a^{8}\\
            &\phantom{\big(}+ 6105246565601165070721815 a^{6}\\
            &\phantom{\big(}- 2678596597802952982813140 a^{4}\\
            &\phantom{\big(}+ 703635822130079540168595 a^{2}\\
            &\phantom{\big(}- 83994803296029834943608
            \big).
    \end{aligned}
\end{gather}
Fig.~\ref{f:NSM14} shows a plot of $G(a)$ in the phase \eqref{eq:phase14}.
It has its minimum at the only root of $G'(a)$ in this phase.
We define
\begin{equation}\label{eq:f14}
    f(a^2) \defeq 3^{-\frac{9}{14}} a^{\frac{67}{7}} G'(a)
\end{equation}
and $v \defeq a^2$ to obtain a polynomial of degree $15$ with rational coefficients,
namely
\begin{equation}\label{eq:f14expanded}
    \begin{aligned}[b]
        f(v) ={}& \frac{1}{2^{24} \cdot 3^{15} \cdot 5^{5} \cdot 7^{6} \cdot 11^{2} \cdot 13}\\
        &\big( - 1018103123715692435 v^{15}\\
        &\phantom{\big(}+ 17231830904837964816 v^{14}\\
        &\phantom{\big(}- 50384128456897156089 v^{13}\\
        &\phantom{\big(}- 1298484357790911467904 v^{12}\\
        &\phantom{\big(}+ 18853052319918661538421 v^{11}\\
        &\phantom{\big(}- 134210060479957033956384 v^{10}\\
        &\phantom{\big(}+ 618613960836724635966183 v^{9}\\
        &\phantom{\big(}- 1974707405188629201419472 v^{8}\\
        &\phantom{\big(}+ 4363698900317368804089183 v^{7}\\
        &\phantom{\big(}- 6221111401101448657037232 v^{6}\\
        &\phantom{\big(}+ 5275022790841652128092813 v^{5}\\
        &\phantom{\big(}- 813004897455917734617504 v^{4}\\
        &\phantom{\big(}- 3663147939360699042433089 v^{3}\\
        &\phantom{\big(}+ 2857169704323149848334016 v^{2}\\
        &\phantom{\big(}- 1078908260599455294925179 v\\
        &\phantom{\big(}+ 167989606592059669887216 \big).
    \end{aligned}
\end{equation}
If $v_0$ denotes the second positive root of $f$, then the minimum of $G(a)$
lies at
\begin{equation}\label{eq:aopt14}
    \aopt = \sqrt{v_0} \approx 1.314\,224\,989\,311,
\end{equation}
leading to an NSM of
\begin{equation}\label{eq:Gopt14}
    G(\aopt) \approx 0.069\,261\,778\,717.
\end{equation}
The numerically estimated NSM was $0.069\,261 \pm 0.000\,002$ \cite[Tab.~II]{agrell24opt}.

For comparison, the previously lowest reported NSM in $14$ dimensions is
$0.069\,52$ \cite{lyu22}, which is higher than the NSM of \eqref{eq:gen14} for
any $a$ in the phase \eqref{eq:phase14}.

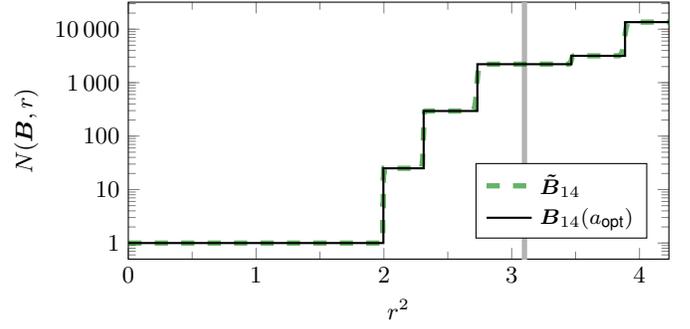
\begin{figure}\centering
    \begin{tikzpicture}
        \sffamily\small
        \begin{axis}[
            width=.99\columnwidth, height=5cm,
            xmin=0, xmax=4.23,
            minor x tick num=1,
            xlabel={$r^2$},
            xlabel near ticks,
            ymode=log,
            log ticks with fixed point,
            ymin=0.5, ymax=32000,
            ylabel={$N(\bB,r)$},
            ylabel near ticks,
            ytick distance=10,
            yticklabel style={/pgf/number format/1000 sep=\,}, 
            legend style={at={(.96,.08)}, anchor=south east, legend cell align=left, font=\footnotesize}
        ]
            \legend{$\tB_{14}$, $\bB_{14}(\aopt)$}
\draw[line width=2, black!30] (3.1,.1) -- (3.1,100000.);
\addplot[green!50!black, opacity=.6, line width=2, dash pattern=on 5pt off 5pt] table{fig-data/logth14.txt};
\addplot[line width=.8] table{
0.	1.
1.99599	1.
1.99599	25.
2.31126	25.
2.31126	295.
2.73144	295.
2.73144	2215.
3.46689	2215.
3.46689	3175.
3.88707	3175.
3.88707	13543.
3.99197	13543.
3.99197	13567.
4.23	13567.
};
        \end{axis}
    \end{tikzpicture}
    \caption{%
        Theta image of the conjectured optimal $14$-dimensional lattice,
        rescaled to determinant 1 and superimposed on
        the theta image of the numerically optimized lattice in Fig.~\ref{f:theta14}.
        The three shells to the left of the gray bar were used
        to determine the parametric lattice from the numerical one.
    }
    \label{f:theta14exact}
\end{figure}

We also calculated the second moment matrix \eqref{eq:Uab} of the lattice \eqref{eq:gen14}
as a function of $a$.
It is
\begin{equation}\label{eq:Uab14}
    \bU =
    \left[
        \begin{array}{cc}
            \alpha(a) \bI_{10} & \bzero \\
            \bzero & \beta(a) \bI_4
        \end{array}
    \right],
\end{equation}
where
\begin{equation}\label{eq:ab14}
    \begin{aligned}[b]
        \alpha(a) &= V(a)^{\frac{1}{7}} G(a) - \frac{5103}{10 V(a)^2} f(a^2) \\
        \beta(a) &= V(a)^{\frac{1}{7}} G(a) + \frac{5103}{4 V(a)^2} f(a^2).
    \end{aligned}
\end{equation}
Recall that a necessary (but not sufficient) condition for the optimality of a
lattice quantizer is that $\bU$ is proportional to the identity matrix.
From \eqref{eq:ab14}, it is clear that $\bU \propto \bI_{14}$ is equivalent to
$f(a^2) = 0$ and thus also equivalent to $G'(a) = 0$.
This condition is satisfied at $\aopt$.
Although the condition only proves local optimality,
we conjecture that $\bB_{14}(\aopt)$ generates the globally optimal
$14$-dimensional lattice quantizer. The exact theta image of the new
lattice is shown in Fig.~\ref{f:theta14exact}.

\subsection{Further properties of the Voronoi region}

The Voronoi region of the lattice \eqref{eq:gen14} in phase \eqref{eq:phase14}
has $13\,542$ facets, $1\,669\,150\,800$ vertices, and overall
$5\,223\,614\,689\,453$
faces in dimensions $0$ through $14$.
Under its automorphism group
the individual number of classes of faces in dimensions $0$ through $14$ is,
respectively,
     $747$, 
  $6\,211$, 
 $24\,871$, 
 $60\,135$, 
 $94\,264$, 
$100\,782$, 
 $76\,698$, 
 $42\,880$, 
 $17\,739$, 
  $5\,345$, 
  $1\,166$, 
     $195$, 
      $31$, 
       $5$, 
and    $1$, 
making in total $431\,070$ equivalence classes of faces across all dimensions.

The lattice has only $24$ shortest nonzero
vectors, which is its \emph{kissing number.}
Its packing radius and density are, respectively, $\rho=a/\sqrt{2}$ and
$\Delta=\pi^7 a^{10}/(5806080\sqrt{3})$.
At $\aopt$, the corresponding numerical values are approximately $0.929\,297$ and $0.004\,616$.
The covering radius and thickness are $\sqrt{(a^2 + 3)(3a^2 + 1)/(6a^2)}$ and
$\sqrt{3}\pi^7(a^2+3)^7(3a^2+1)^7/(38\,093\,690\,880 a^{18})$,
which are approximately $2.819\,748$ and $18.264\,550$ at $\aopt$, respectively.

\section{Dimension 13: Glued $A_7 \times D_5 \times \Z$}
\label{sec:13D}

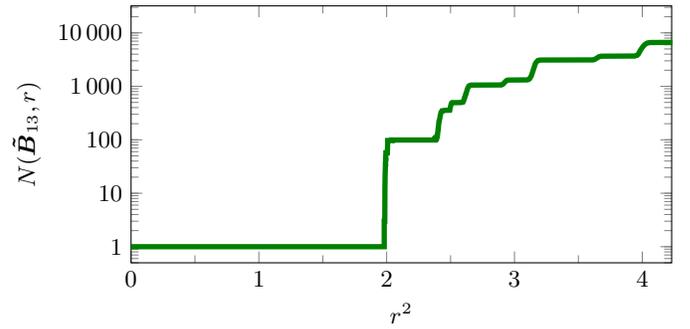
\begin{figure}\centering
    \begin{tikzpicture}
        \sffamily\small
        \begin{axis}[
            width=.99\columnwidth, height=5cm,
            xmin=0, xmax=4.23,
            minor x tick num=1,
            xlabel={$r^2$},
            xlabel near ticks,
            ymode=log,
            log ticks with fixed point,
            ymin=0.5, ymax=32000,
            ylabel={$N(\tB_{13},r)$},
            ylabel near ticks,
            ytick distance=10,
            yticklabel style={/pgf/number format/1000 sep=\,}, 
            legend style={at={(.96,.08)}, anchor=south east, legend cell align=left, font=\footnotesize}
        ]
        \addplot[green!50!black, line width=2] table{fig-data/logth13.txt};
        \end{axis}
    \end{tikzpicture}
    \caption{%
        Theta image of a numerically optimized $13$-dimensional lattice.
    }
    \label{f:theta13}
\end{figure}

\subsection{Identification based on the theta image}

The theta image of the numerical lattice with the lowest NSM
we found in $13$ dimensions is shown in Fig.~\ref{f:theta13}.
When compared with the $14$-dimensional lattice in Fig.~\ref{f:theta14}, the
vertical steps are not as clearly separated from each other.
Nevertheless, the same identification technique \cite{agrell24opt} was applied
and yielded in this case an exact lattice $\gL$, with no unknown parameter.
We identified $\gL$ as glued $A_7 \times D_5 \times \sqrt{2}\Z$, 
with generator
\begin{equation}\label{eq:gen13plain}
    \begin{aligned}[b]
        &\bB_{13}' =\\
    &\left[
        \begin{array}{@{\!\!}*{8}{@{\;}c@{\;}};{2pt/2pt}@{}c@{}*{5}{@{\;}c@{\;}}@{\!\!}}
            -1              & 1             & 0              & 0              & 0              & 0              & 0              & 0              && 0  & 0  & 0  & 0  & 0  \\
            -1              & 0              & 1             & 0              & 0              & 0              & 0              & 0              && 0  & 0  & 0  & 0  & 0  \\
            -1              & 0              & 0              & 1             & 0              & 0              & 0              & 0              && 0  & 0  & 0  & 0  & 0  \\
            -1              & 0              & 0              & 0              & 1             & 0              & 0              & 0              && 0  & 0  & 0  & 0  & 0  \\
            -1              & 0              & 0              & 0              & 0              & 1             & 0              & 0              && 0  & 0  & 0  & 0  & 0  \\
            -1              & 0              & 0              & 0              & 0              & 0              & 1             & 0              && 0  & 0  & 0  & 0  & 0  \\
            \rule[-1.5ex]{0pt}{0pt}%
            {-1}              & 0              & 0              & 0              & 0              & 0              & 0              & 1             && 0  & 0  & 0  & 0  & 0  \\
            \hdashline[2pt/2pt]
            \tstrut
            0              & 0              & 0              & 0              & 0              & 0              & 0              & 0              && 2  & 0  & 0  & 0  & 0  \\
            0              & 0              & 0              & 0              & 0              & 0              & 0              & 0              && 1  & 1  & 0  & 0  & 0  \\
            0              & 0              & 0              & 0              & 0              & 0              & 0              & 0              && 1  & 0  & 1  & 0  & 0  \\
            0              & 0              & 0              & 0              & 0              & 0              & 0              & 0              && 1  & 0  & 0  & 1  & 0  \\
            \rule[-1.5ex]{0pt}{0pt}
            0              & 0              & 0              & 0              & 0              & 0              & 0              & 0              && 1  & 0  & 0  & 0  & 1  \\
            \hdashline[2pt/2pt]
            \tstrut
            \tfrac{11}{16} & \tfrac{11}{16} & \tfrac{11}{16} & -\tfrac{5}{16} & -\tfrac{5}{16} & -\tfrac{5}{16} & -\tfrac{5}{16} & -\tfrac{5}{16} && \h & \h & \h & \h & \h \\
        \end{array}
    \right].
  \end{aligned}
\end{equation}
This is a case of generalized gluing as described in Sec.~\ref{sec:gengluing}.
The scale factor $\sqrt{2}$ serves to give all three component lattices
the same shortest vector norm $\sqrt{2}$.
We defer a description of the gluing structure to the more generic
lattice \eqref{eq:gen13}, of which $\bB_{13}'$ is a special case.

The lattice $\gL$ was analyzed using the methods in
\cite{Pook-Kolb2022Exact, pook-kolb23}.
This was easier than for the $14$-dimensional lattice because there is no
free parameter.
The resulting NSM of
\begin{equation}\label{eq:G13plain}
    \begin{aligned}[b]
        G ={}& \frac{264643025208158502912098205658743146287}{2^{81} \cdot 3^{9} \cdot 5^{3} \cdot 7^{4} \cdot 11^{2} \cdot 13^{3}}\\
            \approx{}& 0.069\,698\,255\,940
    \end{aligned}
\end{equation}
is lower than all previously reported NSMs in $13$ dimensions
(most recently $0.071\,035$ \cite{agrell23}, $0.070\,974$ \cite{agrell24K12},
and $0.069\,901$ \cite{Pook-Kolb2022Exact}).

As in $14$ dimensions, we calculated the second moment matrix \eqref{eq:Uab}
to verify the condition $\bU \propto \bI_{13}$, which must be satisfied if the
lattice is locally optimal.
Surprisingly, it turned out that the second moment matrix is {\em not} proportional to the identity.
Instead,
\begin{equation}\label{eq:Uab13plain}
    \bU =
    \left[
        \begin{array}{@{\,}c@{\;\;}c@{\,}}
            \bW_8  & \bzero \\
            \bzero & \alpha \bI_5
        \end{array}
    \right],
\end{equation}
where $\bW_8$ is an $8 \times 8$ matrix with
$\beta$ on the diagonal and $\gamma$ in all other components.
The values of $\alpha$, $\beta$, and $\gamma$ are
\begin{equation}\label{eq:Uab13plain:defs}
    \begin{alignedat}{2}
        \alpha &= \tfrac{304547502154926541417266582464260350511}{2^{81} \cdot 3^{10} \cdot 5^{4} \cdot 7^{4} \cdot 11^{2} \cdot 13^{2}} &&\approx 0.069\,513, \\
        \beta &= \tfrac{9787631469390979346380560690239381767}{2^{83} \cdot 3^{10} \cdot 5 \cdot 7^{4} \cdot 11^{2} \cdot 13^{2}} &&\approx 0.069\,814, \\
        \gamma &= - \tfrac{972574414727556817448919098411598577}{2^{83} \cdot 3^{10} \cdot 5^{4} \cdot 7^{4} \cdot 11^{2} \cdot 13^{2}} &&\approx -0.000\,055.
    \end{alignedat}
\end{equation}
This proves that $\gL$ generated by \eqref{eq:gen13plain} is {\em not} the optimal $13$-dimensional lattice quantizer
and the identification method in \cite{agrell24opt} had failed for the first time.
However, the fact that $\alpha-\beta$ and $\gamma$ are near zero
suggests that a local optimum might be nearby.

\subsection{Optimization of the NSM}
\label{sub:opt13}

To find the local optimum near $\gL$, we apply the algorithm of Sec.~\ref{sec:optimization}.
The linearized perturbation matrix as a function of the step size $\epsilon$
is $\bA_\epsilon = \bI_n-\epsilon\bbU$, with $\bbU$ given by \eqref{eq:Utraceless}
and $\bU$ by \eqref{eq:Uab13plain}. The perturbed generator matrix is
\begin{align}
\bB_{13}' \bA_\epsilon
  &= \bB_{13}' \left(\bI_n - \epsilon \left(\bU - \frac{\tr\bU}{n} \bI_n \right) \right) \\
  &= \left(1 + \frac{\epsilon\tr\bU}{n} \right) \bB_{13}' -\epsilon\bB_{13}'\bU \\
  &= \left(1+\frac{\epsilon}{13}\left(5\alpha+8\beta\right)\right)\bB_{13}' - \epsilon \bB_{13}' \bU
  \label{eq:gen13perturbed}
\end{align}
and the perturbed second moment matrix is
\begin{align}
\bA_\epsilon^T \bU \bA_\epsilon
  &= \bU-\epsilon(\bbU\bU+\bU\bbU) + \epsilon^2\bbU\bU\bbU \label{eq:AUA1} \\
  &= \begin{bmatrix}
    \bW'_8 & \bzero \\
    \bzero & \alpha' \bI_5
  \end{bmatrix} + O(\epsilon^2),
\end{align}
where $\bW'_8$ has $\beta'$ in the diagonal and $\gamma'$ in all other components.
The values of these components as functions of $\epsilon$ are
\begin{align}
  \alpha' &= \alpha + \frac{16\alpha(\beta-\alpha)}{13}\epsilon, \label{eq:alpha} \\
  \beta' &= \beta - \frac{2(5\beta^2+91\gamma^2-5\alpha\beta)}{13}\epsilon, \label{eq:beta} \\
  \gamma' &= \gamma+\frac{2\gamma(5\alpha-18\beta-78\gamma)}{13}\epsilon, \label{eq:gamma}
\end{align}
which can be verified by inserting \eqref{eq:Uab13plain}
(with $\bW_8$ written in terms of $\beta$ and $\gamma$) and
\eqref{eq:Utraceless} into \eqref{eq:AUA1} and
keeping the terms up to linear order in $\epsilon$.

We choose $\epsilon$ so that
$\bA_\epsilon^T \bU \bA_\epsilon \propto \bI_{13}$ is approximately
satisfied.
Imposing $\gamma' = 0$ in \eqref{eq:gamma} yields
\begin{align}
  \epsilon = \epsilon_1 \defeq \frac{13}{2(-5\alpha+18\beta+78\gamma)} \approx 7.1843,
\end{align}
whereas imposing $\alpha' = \beta'$ using \eqref{eq:alpha}--\eqref{eq:beta} yields
\begin{align}
  \epsilon = \epsilon_2 \defeq
    \frac{13(\beta-\alpha)}{2(-8\alpha^2+3\alpha\beta+5\beta^2+91\gamma^2)} \approx 7.1735.
\end{align}
Because $\bU$ \eqref{eq:Uab13plain} is very close to the identity, $\bbU$ is
very close to zero. Therefore, even with $\epsilon\approx 7$,
all entries in $\epsilon\bbU$ are less than $0.0014$.
That $\epsilon_1 \ne \epsilon_2$ means that the second moment matrix
$\bA_\epsilon^T \bU \bA_\epsilon$
of the deformed Voronoi region $\Omega\bA_\epsilon$
will not be proportional to $\bI_{13}$ for any $\epsilon$.
Hence, the multiplication with $\bA_\epsilon$ likely cannot bring the lattice
directly into a local minimum for any $\epsilon$.
Nevertheless, it is remarkable that both $\epsilon_1$
and $\epsilon_2$ satisfy $\gamma'\approx0$ and $\alpha'\approx\beta'$.
This confirms that $\bA_\epsilon$ deforms the Voronoi region to a
good approximation in the direction of a second moment matrix
proportional to $\bI_{13}$,
which is hopefully sufficient to identify the optimal phase.

We will now show that the perturbation using $\bA_\epsilon$ generates a
\emph{generalized glued lattice,} as defined
in Sec.~\ref{sec:gengluing}.
To this end, we generalize $\bB_{13}'$
by gluing $\Lp = a_1 A_7 \times a_2 D_5 \times a_3\sqrt{2}\Z$ for arbitrary positive
scale factors $\ba=[a_1,a_2,a_3]$, instead of gluing
$A_7 \times D_5 \times \sqrt{2}\Z$ as in \eqref{eq:gen13plain}.
This yields
{\setlength{\arraycolsep}{.17em} 
\begin{align} \label{eq:gen13}
&\bB_{13}(\ba) = \notag\\
  &\left[
  \begin{array}{@{\!\!}*{8}{@{\;}c@{\;}};{2pt/2pt}@{}c@{}*{5}{@{\;}c@{\;}}@{\!\!}}
    -a_1                    & a_1 & 0   & 0   & 0   & 0   & 0   & 0   && 0              & 0              & 0              & 0              & 0              \\
    -a_1                    & 0   & a_1 & 0   & 0   & 0   & 0   & 0   && 0              & 0              & 0              & 0              & 0              \\
    -a_1                    & 0   & 0   & a_1 & 0   & 0   & 0   & 0   && 0              & 0              & 0              & 0              & 0              \\
    -a_1                    & 0   & 0   & 0   & a_1 & 0   & 0   & 0   && 0              & 0              & 0              & 0              & 0              \\
    -a_1                    & 0   & 0   & 0   & 0   & a_1 & 0   & 0   && 0              & 0              & 0              & 0              & 0              \\
    -a_1                    & 0   & 0   & 0   & 0   & 0   & a_1 & 0   && 0              & 0              & 0              & 0              & 0              \\
    \rule[-1.5ex]{0pt}{0pt}
    {-a_1}                  & 0   & 0   & 0   & 0   & 0   & 0   & a_1 && 0              & 0              & 0              & 0              & 0              \\
    \hdashline[2pt/2pt]
    \tstrut
    0                       & 0   & 0   & 0   & 0   & 0   & 0   & 0   && 2a_2           & 0              & 0              & 0              & 0              \\
    0                       & 0   & 0   & 0   & 0   & 0   & 0   & 0   && a_2            & a_2            & 0              & 0              & 0              \\
    0                       & 0   & 0   & 0   & 0   & 0   & 0   & 0   && a_2            & 0              & a_2            & 0              & 0              \\
    0                       & 0   & 0   & 0   & 0   & 0   & 0   & 0   && a_2            & 0              & 0              & a_2            & 0              \\
    \rule[-1.5ex]{0pt}{0pt}
    0                       & 0   & 0   & 0   & 0   & 0   & 0   & 0   && a_2            & 0              & 0              & 0              & a_2            \\
    \hdashline[2pt/2pt]
    \tstrut
    b                       & b   & b   & b'  & b'  & b'  & b'  & b'  && \tfrac{a_2}{2} & \tfrac{a_2}{2} & \tfrac{a_2}{2} & \tfrac{a_2}{2} & \tfrac{a_2}{2}
  \end{array}
  \right],
\end{align}}%
where $b \defeq (10a_1+a_3)/16$ and $b' \defeq b-a_1$.

The first seven rows of $\bB_{13}(\ba)$ generate $a_1 A_7$ (embedded in $\R^8$)
and the next five rows generate $a_2 D_5$.
The third component lattice $a_3 \sqrt{2}\Z$ consists of all multiples of
$a_3 [1\;1\;1\;1\;1\;1\;1\;1\;0\;0\;0\;0\;0]/2$ and is orthogonal to the first twelve rows.
That this vector belongs to the lattice can be seen by taking $8$ times the last
row of \eqref{eq:gen13} and adding or subtracting suitable multiples of the other rows.
Indeed, if the last row of \eqref{eq:gen13} would be replaced by
$a_3 [1\;1\;1\;1\;1\;1\;1\;1\;0\;0\;0\;0\;0]/2$, then
the resulting matrix would generate the (unglued) product lattice
$\Lp$.

Constructed from the product lattice, the glued lattice $\gL$
is given by \eqref{eq:gluing} with $\LL=\Lp$ and
the glue group $\Gamma = \{i \bg_1 \colon i=0,\ldots,7\}$,
where $\bg_1$ is the last row of \eqref{eq:gen13}.
Interestingly, $\Gamma$ is not a subgroup of $\Lp^*/\Lp$,
not even for $a_1=a_2=a_3=1$.

The glued lattice $\gL$, like its constituent product lattice $\Lp$,
has $56$ vectors of length $a_1\sqrt{2}$, belonging to $a_1 A_7$,
$40$ vectors of length $a_2\sqrt{2}$, belonging to $a_2 D_5$,
and $2$ vectors of length $a_3\sqrt{2}$, belonging to $a_3\sqrt{2}\Z$.
The shortest nonzero vectors have length $\min(a_1, a_2, a_3)\sqrt{2}$, and so the
number of those vectors depends on the parameter values.
The kissing number
can thus be $98$ (for $a_1 = a_2 = a_3$),
$96$, $58$, $56$, $42$, $40$, or $2$ (for $a_3 < \min(a_1, a_2)$).

As a special case, $\bB_{13}'$ in \eqref{eq:gen13plain} is obtained by setting
$\ba = [1,1,1]$ in \eqref{eq:gen13}.
Another special case of \eqref{eq:gen13} is $\bB_{13}' \bA_\epsilon$ in \eqref{eq:gen13perturbed}
for any $\epsilon$, which can be verified by setting
\begin{align}
a_1 &= 1-\frac{5\beta-5\alpha-13\gamma}{13}\epsilon \label{eq:a1}\\
a_2 &= 1+\frac{8\beta-8\alpha}{13}         \epsilon \\
a_3 &= 1-\frac{5\beta-5\alpha+91\gamma}{13}\epsilon \label{eq:a3}
\end{align}
in \eqref{eq:gen13} and comparing the result
component-wise with \eqref{eq:gen13perturbed}.
We therefore conjecture that the optimal $13$-dimensional
lattice has the form \eqref{eq:gen13} for some $\ba$.
We do not constrain $\ba$ to the single degree
of freedom given by \eqref{eq:a1}--\eqref{eq:a3}, because of the approximations
involved in deriving those relations.
On the other hand, it is unnecessary to consider the whole three-dimensional
parameter space $\ba=[a_1,a_2,a_3]$, because rescaling the lattice by a nonzero
constant $c$ does not affect the NSM \eqref{eq:NSM}
and $c\bB_{13}(\ba) = \bB_{13}(c\ba)$.
We therefore fix $a_1=1$ without loss of generality.

The two-parameter family of lattices \eqref{eq:gen13} with $\ba=[1,a_2,a_3]$
is much more complicated
to analyze than the one-parameter family \eqref{eq:gen14}.
The main reason is that instead of intervals, phases of given Voronoi topology
are now regions in the two-dimensional parameter space.
A second reason is that the symbolic expressions in the calculations for the
NSM and second moment matrix become more complicated due to the additional
variable.

\begin{figure}\centering
    \includegraphics[width=\linewidth]{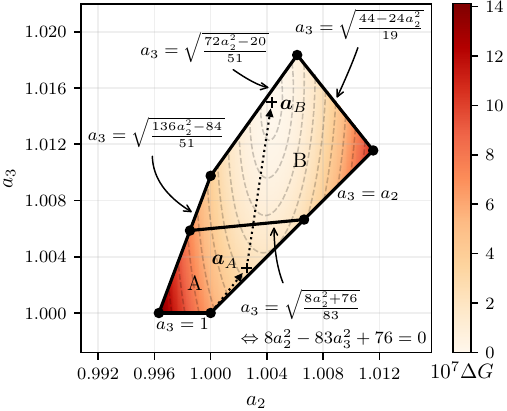}
    \caption{%
        Phases A and B of the lattice \eqref{eq:gen13}.
        Solid lines show phase boundaries where $1$-face lengths vanish and
        the dots mark the points where lines of vanishing $1$-face lengths
        intersect.
        The color gradient indicates the value of
        $\Delta G(\ba) \defeq G(\ba) - G(\baopt)$, where
        $G = G_A$ in phase A, $G = G_B$ in phase B,
        and $G(\baopt)$ is given in \eqref{eq:Gopt13}.
        The functions $G_B$ and $G_A$ are extremely similar throughout the colored
        region (see \eqref{eq:NSM13B} and \eqref{eq:DG}).
        Dashed lines show contours of equal $G$.
        The dotted arrows indicate the search path described in Sec.~\ref{sub:opt13}.
        The minimum of $G$ lies very close to $\baoptA$ (see \eqref{eq:aA_vs_aB}).
    }
    \label{f:phases13AB}
\end{figure}

We analyze the parametric lattice first at the point
$\ba_A = [1,a_2^A, a_3^A]$, where
\begin{equation}\label{eq:phaseApoint}
    \begin{alignedat}[b]{2}
        a_2^A &= 392/391 &&\approx 1.002\,558, \\
        a_3^A &= 314/313 &&\approx 1.003\,195.
    \end{alignedat}
\end{equation}
These values were obtained by setting $\epsilon = \epsilon_1$ in
\eqref{eq:a1}--\eqref{eq:a3}, normalizing $\ba$ by $a_1$, and using
rational approximations to facilitate symbolic calculations.
We call the phase that $\ba_A$ lies in {\em phase A}.
Fig.~\ref{f:phases13AB} shows $\ba_A$ and the phase boundaries identified by
vanishing lengths of $1$-faces.
The additional conditions
($i$)--($iv$) of Sec.~\ref{sec:voronoi} are verified as well.
Note that not all phase boundaries are straight lines, as can be seen from the
expressions defining the boundaries in Fig.~\ref{f:phases13AB}.

The NSM $G_A$ of phase A as a function of $\ba$ is given in Appendix~\ref{app:13A}.
To locate a minimum of $G_A$, we use
$G^A_2 \defeq \partial G_A/\partial a_2$
and
$G^A_3 \defeq \partial G_A/\partial a_3$.
A local minimum of $G_A$ in $\ba$ has $G^A_2 = G^A_3 = 0$.
However, no such point exists within phase A.
We therefore need to repeat the analysis for
another phase.

According to the iterative strategy outlined in Sec.~\ref{sec:optimization},
the next point to analyze is the local minimum of $G_A(\ba)$
closest to (with respect to the Euclidean distance in parameter space)
phase A.
It is located at $\baoptA \approx [1, 1.004\,336, 1.014\,983]$.%
\footnote{%
    During this work, we analyzed the lattice at $[1, 1.004, 1.008]$
    instead of $\baoptA$ and only later realized that $\baoptA$ would have been a
    better choice.
    Happily, both points are in the same phase and thus yield
    the same expression for $G_B$.
}
Note that $G_A(\baoptA)$ is not the NSM of $\bB_{13}(\baoptA)$, but it is the best
approximation available at this point of the analysis.
The phase determined from constructing the Voronoi region at
$\baoptA$ is also shown in Fig.~\ref{f:phases13AB}, with exact expressions
for the boundaries given in the figure.
We call this {\em phase B}.
The criteria ($i$)--($iv$) of Sec.~\ref{sec:voronoi} are satisfied
within both phases.

The NSM in phase B is, still assuming $a_1=1$,
\begin{equation}\label{eq:NSM13B}
    G_B(\ba) = G_A(\ba) - \frac{\left(8 a_{2}^{2} - 83 a_{3}^{2} + 76\right)^{14}}{2^{78} \cdot 3^{5} \cdot 5^{3} \cdot 7^{2} \cdot 11 \cdot 13^{2}\ V(\ba)^{2+2/13}}
    ,%
    \rule[-2.2ex]{0pt}{0pt}
\end{equation}
where $V(\ba) = \lvert\det \bB_{13}(\ba)\rvert = a_2^5 a_3$.
Therefore, the NSM $G(\ba)$ of \eqref{eq:gen13}, which equals $G_A(\ba)$ in phase A
and $G_B(\ba)$ in phase B, is $13$ (but not $14$) times differentiable
on the boundary between the two phases, which is given by $8a_2^2-83a_3^2+76=0$.
This observation demonstrates that $G_A$ is a good approximation of $G$
also outside phase A, which in turn justifies the strategy of taking the minimum of $G_A$ as
the next point to analyze also when it lies outside phase A.

The minimum of $G_B(\ba)$ is located at the only common root of
$G^B_2 \defeq \partial G_B/\partial a_2$
and
$G^B_3 \defeq \partial G_B/\partial a_3$ in the interior of phase B.
The two equations $G^B_2 = 0$ and $G^B_3 = 0$ are equivalent to
$f^B_2 = 0$ and $f^B_3 = 0$, where $f^B_i$ are rational polynomials of
degree $14$ in $v_2 \defeq a_2^2$ and $v_3 \defeq a_3^2$.
They are obtained via
\begin{equation}\label{eq:fB}
    \begin{aligned}[b]
        f^B_2(a_2^2, a_3^2) &\defeq a_2^{\frac{153}{13}} a_3^{\frac{28}{13}} G^B_2(\ba), \\
        f^B_3(a_2^2, a_3^2) &\defeq a_2^{\frac{140}{13}} a_3^{\frac{41}{13}} G^B_3(\ba).
    \end{aligned}
\end{equation}
Explicit expressions for $f^B_2$ and $f^B_3$ are given in the online
supplementary material \cite{ancillary-files}.

The minimum of $G_B$ in phase B is shown in Fig.~\ref{f:phases13AB}.
Its value is
\begin{equation}\label{eq:Gopt13}
    G(\baopt) = G_B(\baopt) \approx 0.069\,697\,638\,992,
\end{equation}
where
\begin{equation}\label{eq:aopt13}
    \begin{aligned}[b]
        \baopt \approx [1,\ 1.004\,336\,185\,575,\ 1.014\,983\,466\,336].
    \end{aligned}
\end{equation}
The minimum \eqref{eq:Gopt13} is not far from the numerically estimated NSM
$0.069\,696 \pm 0.000\,002$ \cite[Tab.~II]{agrell24opt}.
This is slightly lower than the NSM \eqref{eq:G13plain} of the same parametric lattice
$\bB'_{13} = \bB_{13}(1,1,1)$.

Interestingly, the difference between $G_A$ and $G_B$ at $\baopt$ is much
smaller than the range of variation of the NSM across phases A and B,
\begin{equation}\label{eq:DG}
    G_B(\baopt) - G_A(\baopt) \approx 1.8 \cdot 10^{-31}.
\end{equation}
Furthermore, the minima of $G_A$ and $G_B$ lie very close to
each other, with a difference in parameters of only
\begin{equation}\label{eq:aA_vs_aB}
    \baoptA - \baopt \approx [0,\ 4 \cdot 10^{-27},\ 9 \cdot 10^{-26}].
\end{equation}
This further supports the assumption that the NSM behaves very similarly
in neighboring phases.
Based on this and the fact that the initial numerical search converged near
$\baopt$, we conjecture that \eqref{eq:Gopt13} is the minimum NSM across all
phases and hence that $\bB_{13}(\baopt)$ generates
the globally optimal lattice quantizer in $13$ dimensions.

As in the $14$-dimensional case, we also calculated the second moment matrix
\eqref{eq:Uab} of \eqref{eq:gen13} in phase B as a function of $\ba$.
As expected from the result \eqref{eq:Uab13plain} for the nonparametric
lattice \eqref{eq:gen13plain}, the second moment matrix is again of the form
\begin{equation}\label{eq:name}
    \bU =
    \left[
        \begin{array}{@{\,}c@{\;\;}c@{\,}}
            \bW_8  & \bzero \\
            \bzero & \alpha(\ba)\bI_5
        \end{array}
    \right],
\end{equation}
where $\bW_8$ is an $8 \times 8$ matrix with
$\beta(\ba)$ on the diagonal and $\gamma(\ba)$ in all other components.
The functions
$\alpha(\ba)$,
$\beta(\ba)$, and
$\gamma(\ba)$
can be expressed as
\begin{equation}\label{eq:Uab13defs}
    \begin{aligned}[b]
        \alpha(\ba) &= V(\ba)^{\frac{2}{13}} G^B(\ba) + \frac{13}{10 V(\ba)^2} f^B_2(a_2^2, a_3^2), \\
        \beta(\ba) &= V(\ba)^{\frac{2}{13}} G^B(\ba) - \frac{13}{16 V(\ba)^2} f^B_2(a_2^2, a_3^2), \\
        \gamma(\ba) &= \frac{13}{112 V(\ba)^2} \left[ f^B_2(a_2^2, a_3^2) + 8 f_3^B(a_2^2, a_3^2) \right].
    \end{aligned}
\end{equation}
The condition $\bU \propto \bI_{13}$ is equivalent to
$\alpha(\ba) = \beta(\ba)$ and $\gamma(\ba) = 0$,
which in turn is equivalent to
$f^B_2(a_2^2, a_3^2) = f^B_3(a_2^2, a_3^2) = 0$.
Since $f^B_2$ and $f^B_3$ both vanish at $\baopt$, the necessary condition
$\bU \propto \bI_{13}$ for the lattice to be a
locally optimal lattice quantizer is satisfied.

\subsection{Further properties of the Voronoi region}

The automorphism group $\Aut(\LL)$ of this lattice for generic values of $a_2$ and $a_3$
has order $154\,828\,800$.
In terms of the lattice vectors $\bu\bB_{13}(\ba)$, where
$\bB_{13}(\ba)$ is given in \eqref{eq:gen13} and $\bu\in\Z^{13}$,
the group consists of 
all permutations of the first five components,
all permutations of the last eight components,
an even number of sign changes among the first five components,
and, of course, an overall sign change.

\begin{table*}\centering
    \caption{Number of faces and face classes of the lattice \eqref{eq:gen13} at
    $a_2=a_3=1$ compared with phases A and B.}
    \label{t:13faces}
    \begin{tabular}{*{7}r}
        \hline
        & \multicolumn{2}{c}{$a_2=a_3=1$}
        & \multicolumn{2}{c}{phase A}
        & \multicolumn{2}{c}{phase B}
        \\
        dim.
        & \multicolumn{1}{c}{faces} & \multicolumn{1}{c}{classes}
        & \multicolumn{1}{c}{faces} & \multicolumn{1}{c}{classes}
        & \multicolumn{1}{c}{faces} & \multicolumn{1}{c}{classes}
        \\
        \hline\hline
                 $0$ &      $913\,062\,580$ &   $2\,380$ &      $961\,419\,700$ &   $2\,429$ &      $960\,828\,340$ &   $2\,444$ \\
                 $1$ &   $8\,323\,478\,208$ &  $10\,402$ &   $8\,510\,132\,928$ &  $10\,561$ &   $8\,504\,452\,288$ &  $10\,606$ \\
                 $2$ &  $31\,111\,318\,496$ &  $21\,388$ &  $31\,451\,395\,296$ &  $21\,636$ &  $31\,432\,418\,016$ &  $21\,693$ \\
                 $3$ &  $64\,038\,899\,360$ &  $29\,081$ &  $64\,426\,365\,600$ &  $29\,334$ &  $64\,395\,704\,480$ &  $29\,373$ \\
                 $4$ &  $81\,053\,843\,332$ &  $29\,948$ &  $81\,353\,017\,732$ &  $30\,139$ &  $81\,326\,693\,252$ &  $30\,148$ \\
                 $5$ &  $66\,158\,634\,484$ &  $24\,433$ &  $66\,317\,190\,644$ &  $24\,550$ &  $66\,305\,273\,844$ &  $24\,546$ \\
                 $6$ &  $35\,348\,681\,664$ &  $16\,073$ &  $35\,404\,233\,664$ &  $16\,130$ &  $35\,401\,671\,104$ &  $16\,122$ \\
                 $7$ &  $12\,255\,202\,196$ &   $8\,571$ &  $12\,266\,760\,596$ &   $8\,590$ &  $12\,266\,563\,476$ &   $8\,586$ \\
                 $8$ &   $2\,670\,432\,810$ &   $3\,691$ &   $2\,671\,508\,010$ &   $3\,694$ &   $2\,671\,508\,010$ &   $3\,694$ \\
                 $9$ &      $343\,917\,632$ &   $1\,266$ &      $343\,917\,632$ &   $1\,266$ &      $343\,917\,632$ &   $1\,266$ \\
                $10$ &       $23\,483\,364$ &      $343$ &       $23\,483\,364$ &      $343$ &       $23\,483\,364$ &      $343$ \\
                $11$ &           $695\,818$ &       $71$ &           $695\,818$ &       $71$ &           $695\,818$ &       $71$ \\
                $12$ &             $5\,454$ &       $10$ &             $5\,454$ &       $10$ &             $5\,454$ &       $10$ \\
                $13$ &                  $1$ &        $1$ &                  $1$ &        $1$ &                  $1$ &        $1$ \\
        \hline total & $302\,241\,655\,399$ & $147\,658$ & $303\,730\,126\,439$ & $148\,754$ & $303\,633\,215\,079$ & $148\,903$ \\
        \hline
    \end{tabular}
\end{table*}

The Voronoi region of this lattice in phase B has
$960\,828\,340$ vertices, $5\,454$ facets, and in total
$303\,633\,215\,079$ faces across all dimensions from $0$ to $13$.
Tab.~\ref{t:13faces} lists these values as well as the number of faces
and face classes in dimensions $0$ through $13$ for $\bB'_{13}$
and $\bB_{13}(\ba)$ in phases A and B.

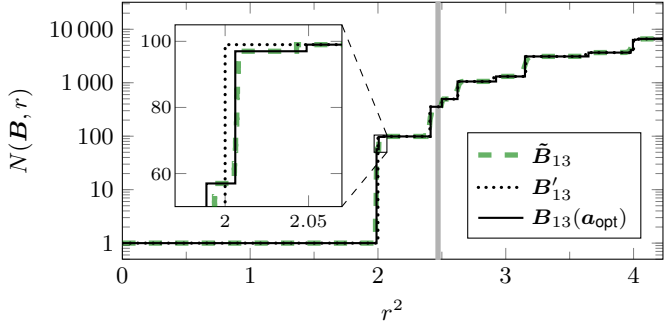
\begin{figure}\centering
    \begin{tikzpicture}
        \sffamily\small
        \begin{axis}[
            width=.99\columnwidth, height=5cm,
            xmin=0, xmax=4.23,
            minor x tick num=1,
            xlabel={$r^2$},
            xlabel near ticks,
            ymode=log,
            log ticks with fixed point,
            ymin=0.5, ymax=32000,
            ylabel={$N(\bB,r)$},
            ylabel near ticks,
            ytick distance=10,
            yticklabel style={/pgf/number format/1000 sep=\,}, 
            legend style={at={(.96,.08)}, anchor=south east, legend cell align=left, font=\footnotesize}
        ]
            \legend{$\tB_{13}$, $\bB'_{13}$, $\bB_{13}(\baopt)$}
            \draw[line width=2, black!30] (2.47,.1) -- (2.47,100000.);
            \addplot[green!50!black, opacity=.6, line width=2, dash pattern=on 5pt off 5pt] table{fig-data/logth13.txt};
            \addplot[line width=1.2, dash pattern=on 0pt off 2.8pt, line cap=round] table{ 
0.	1.
2.	1.
2.	99.
2.40625	99.
2.40625	355.
2.5	355.
2.5	495.
2.625	495.
2.625	1055.
2.90625	1055.
2.90625	1311.
3.15625	1311.
3.15625	3103.
3.625	3103.
3.625	3663.
4.	3663.
4.	6605.
4.23	6605.
            };
            \addplot[line width=.8] table{
0.	1.
1.9888	1.
1.9888	57.
2.00608	57.
2.00608	97.
2.04884	97.
2.04884	99.
2.41202	99.
2.41202	355.
2.50101	355.
2.50101	495.
2.62269	495.
2.62269	1055.
2.92423	1055.
2.92423	1311.
3.15031	1311.
3.15031	3103.
3.64712	3103.
3.64712	3663.
3.9776	3663.
3.9776	4083.
3.99488	4083.
3.99488	6323.
4.01217	6323.
4.01217	6413.
4.03764	6413.
4.03764	6525.
4.05493	6525.
4.05493	6605.
4.23	6605.
            };
        \coordinate (sw) at (axis cs:1.97,50);
        \coordinate (se) at (axis cs:2.07,50);
        \coordinate (ne) at (axis cs:2.07,105);
        \draw (sw) rectangle (ne); 
        \end{axis}
        \begin{axis}[ 
            name=inset,
            width=3.8cm, height=4.0cm,
            xshift=.7cm,yshift=.7cm,
            xmin=1.97, xmax=2.07,
            xticklabel style={font=\scriptsize},
            xtick distance=.05,
            xminorticks=false,
            ymin=50, ymax=105,
            ylabel near ticks,
            yminorticks=false,
            yticklabel style={font=\scriptsize,xshift=2pt}
            ]
            \addplot[green!50!black, opacity=.6, line width=2, dash pattern=on 5pt off 5pt] table{fig-data/th13inset.txt};
            \addplot[line width=1.2, dash pattern=on 0pt off 2.8pt, line cap=round] table{ 
0.	1.
2.	1.
2.	99.
2.40625	99.
            };
            \addplot[line width=.8] table{
0.	1.
1.9888	1.
1.9888	57.
2.00608	57.
2.00608	97.
2.04884	97.
2.04884	99.
2.41202	99.
            };
          \end{axis}
        \draw [dashed] (se) -- (inset.south east);        
        \draw [dashed] (ne) -- (inset.north east);        
    \end{tikzpicture}
    \caption{%
        Theta image of the suboptimal $\bB_{13}'$ and
        the conjectured optimal $\bB_{13}(\baopt)$,
        both rescaled to determinant 1 and superimposed on
        the theta image of the numerically optimized lattice in Fig.~\ref{f:theta13}.
        The inset is in linear scale to better visualize the three steps at $r^2=1.99$, $2.01$, and $2.05$.
        The two shells to the left of the gray bar were used
        to determine $\bB_{13}'$.
    }
    \label{f:theta13exact}
\end{figure}

The theta images of $\bB'_{13}$ and $\bB_{13}(\baopt)$ are shown in Fig.~\ref{f:theta13exact}.
The $98$ shortest nonzero vectors of $\bB'_{13}$, which were mentioned after \eqref{eq:gen13plain},
split into three classes of vectors in $\bB_{13}(\ba)$, with squared norms
$2a_1^2$, $2a_2^2$, and $2a_3^2$, and orbit lengths $56$, $40$, and $2$, respectively.
The step from $1$ to $99$ at $r^2 = 2$ in the theta image of $\bB'_{13}$ corresponds to
three stacked steps for $\bB_{13}(\baopt)$, at $r^2=1.99$, $2.01$, and $2.05$, respectively.
This explains why the lattice identification method in \cite{agrell24opt} failed for $\tB_{13}$.
The method cannot distinguish between a single lattice shell with numerical fluctuations
and several closely-spaced shells.

Some properties of this parametric lattice change nontrivially even within a
phase, i.e., without occurring together with a change of the topology of
$\Omega$.
For example, the kissing number in both phases A and B is
\begin{equation}\label{eq:kissing13}
    \tau = \begin{cases}
        56 & a_2 > 1,\\
        96 & a_2 = 1,\\
        40 & a_2 < 1,\\
    \end{cases}
\end{equation}
still assuming $a_1=1$. (It is $2$, $40$, $42$, $56$, $58$, and $98$ in
different phases just south of phase A, i.e., for $a_3 \le \max(1, a_2)$.)
Accordingly, the packing radius in phases A and B is
$\rho=\min(1,a_2)/\sqrt{2}$
with corresponding density
\begin{align}
\Delta = \frac{128 \pi^{6} \rho^{13}}{135135 a_{2}^{5} a_{3}}.
\end{align}
At $\baopt$ in phase B, we get
$\rho \approx 0.707\,107$ and $\Delta \approx 0.009\,700$.

The covering radius $R$ and thickness $\Theta$ in both phases are
\begin{equation}\label{eq:R_and_Th}
    \begin{aligned}[b]
        R &= \frac{\sqrt{2} \sqrt{64 a_{2}^{4} + 48 a_{2}^{2} a_{3}^{2} - 64 a_{2}^{2} + 9 a_{3}^{4} - 8 a_{3}^{2} + 144}}{16},\\
        \Theta &= \frac{128 \pi^{6} R^{13}}{135135 a_{2}^{5} a_{3}}
    \end{aligned}
\end{equation}
Their respective numerical values at $\baopt$ in phase B are approximately
$1.236\,648$ and $13.889\,470$.

\subsection{Exploring the phase structure}

\begin{figure}\centering
    \includegraphics[width=\linewidth]{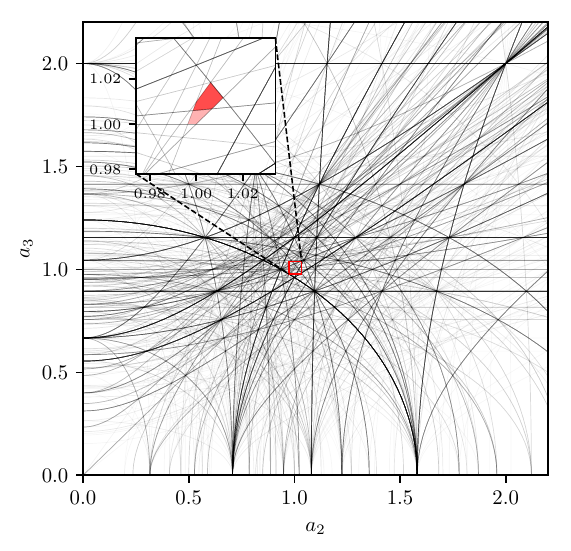}
    \caption{%
        Lines where lengths of $1$-faces in phase B become zero.
        The lighter and darker shaded regions in the inset show
        phases A and B from Fig.~\ref{f:phases13AB}, respectively.
        Note that these lines may not accurately capture the phase
        transitions of $\bB_{13}(\ba)$ outside phases A and B.
    }
    \label{f:phases13busy}
\end{figure}

Phases A and B of the lattice, shown in Fig.~\ref{f:phases13AB}, already show
an intricate geometry, with some boundaries consisting of curved lines.
One can gain an insight into the neighboring phases by looking for lines of
vanishing $1$-face lengths in the face hierarchy of a certain phase
evaluated at parameters {\em outside} its phase.
If those $1$-faces are still part of the Voronoi region, then their vanishing
does indeed indicate a phase boundary.
In fact, phase B can be completely characterized by vanishing lengths of $1$-faces
in phase A and vice versa.
Fig.~\ref{f:phases13busy} shows the curves along which lengths of $1$-faces of the
Voronoi region in phase B become zero.
Semitransparent curves are drawn for each class of $1$-faces, so that
darker curves indicate a larger number of $1$-face classes with vanishing
lengths.
These results should be interpreted as suggestive of
the complexity of the {\em phase landscape.}
More robust pictures can be made by constructing the
Voronoi region for multiple points in the $(a_2, a_3)$-plane and comparing the
resulting phase landscapes.

\section{Conclusions}
\label{sec:conclusions}

By combining several analysis methods,
we obtain new lattice quantizers
in dimensions $13$ and $14$, which we conjecture are optimal,
based on numerical and analytic evidence.
The key steps are to first find an approximate lattice using
stochastic gradient descent.
This is followed by the exact identification as a
generalized glued product lattice.
The second moment matrix then provides a clue whether further refinement is
necessary, which was an important step in the analysis of the
$13$-dimensional lattice.
Parameterizing the product lattice by scaling the components has proved
to be a successful route.
The optimal parameters are determined via the common root of two polynomials
of degree $14$ (in dimension $13$)
and a root of a single polynomial of degree $15$ (in dimension $14$).

A somewhat surprising result is that
the optimal parameters in $13$ dimensions are close but not equal to unity.
At exactly $1$, the lattice is not even locally optimal, since its
second moment matrix is not proportional to the identity.
The difference is insignificant in practice but demonstrates
the elusive nature of optimal lattice quantizers.
Furthermore,
by visualizing the points in parameter space where lengths of $1$-faces
become zero (Fig.~\ref{f:phases13busy}), we found that Voronoi regions
can undergo a much higher number of topological changes than we initially
expected.

We conjecture that the optimal $13$-dimensional lattice quantizer is
generated by $\bB_{13}(1, \sqrt{v_{2,0}}, \sqrt{v_{3,0}})$ in \eqref{eq:gen13},
where $[v_{2,0}, v_{3,0}]$ is the common root of the polynomials $f^B_2$ and
$f^B_3$ closest to $[1,1]$. The full expressions of $f^B_2$ and $f^B_3$ are
given in the online supplementary material \cite{ancillary-files}.
We also conjecture that the optimal $14$-dimensional lattice quantizer is
generated by $\bB_{14}(\sqrt{v_0})$ in \eqref{eq:gen14},
where $v_0$ is the second positive root of $f(v)$ in \eqref{eq:f14expanded}.

Together with the conjectured optimal lattices
in dimensions below $13$ and in dimension $16$
\cite[Chap.~1]{conway99splag}, 
\cite{allen21}, 
\cite{agrell24opt}, 
we now believe that the list of optimal lattices quantizers
is complete up to dimension $16$.

\bibliography{refs}

\appendices

\section{Exact NSM in dimension $13$}
\label{app:13A}

The NSM of the two-parameter lattice \eqref{eq:gen13} with $a_1=1$ in phase A
(see Fig.~\ref{f:phases13AB}) is
\begin{equation}\label{eq:NSM13A}
    G_A(\ba) = \frac{\sum_{i=0}^{14}\sum_{j=0}^{14-i} c_{i,j}a_2^{2i}a_3^{2j}}{2^{81} \cdot 3^{10} \cdot 5^{4} \cdot 7^{4} \cdot 11^{2} \cdot 13^{3}\ V(\ba)^{2+2/n}}
    ,%
    \rule[-2.2ex]{0pt}{0pt}
\end{equation}
where $V(\ba) = a_2^5 a_3$ and $n = 13$.
The coefficients $c_{i,j}$ are integers given in Tab.~\ref{t:wij}.

\begin{table*}\centering
    \caption{Coefficients of \eqref{eq:NSM13A}.}
    \label{t:wij}
    {%
    \renewcommand\arraystretch{1.2}  
    \begin{tabular}{rrr|rrr}
        \hline
        $i$ & $j$  & \multicolumn{1}{c|}{$c_{i,j}$}              &  $i$ &  $j$ & \multicolumn{1}{c}{$c_{i,j}$} \\ \hline\hline
        $0$ &  $0$ &      $237032097068933436616799735576002560$ &  $4$ &  $6$ &   $1074693958390657454873853548257188249600$\\
        $0$ &  $1$ &    $-1463207626460982855861770593966751744$ &  $4$ &  $7$ &    $285838925005138976162976386866898534400$\\
        $0$ &  $2$ &     $5934898857208683473779516470997286912$ &  $4$ &  $8$ &    $102338181627589302426266247730490572800$\\
        $0$ &  $3$ &    $-6206764416204081335945257654983589888$ &  $4$ &  $9$ &     $34784275308690432469536266656562380800$\\
        $0$ &  $4$ &     $3250756445663451686494512850303713280$ &  $4$ & $10$ &      $7176569883812137057957384842426060800$\\
        $0$ &  $5$ &    $27531155233952148807494665749780758528$ &  $5$ &  $0$ &   $-489385102760026409815066861915477114880$\\
        $0$ &  $6$ &   $-52726816487909003990881825349185634304$ &  $5$ &  $1$ &   $-349855410700338542688799497434840432640$\\
        $0$ &  $7$ &    $78104448955066346519712454831909502976$ &  $5$ &  $2$ &  $-4594184246897310463538302471981812940800$\\
        $0$ &  $8$ &   $-69925347164027747089931918702597099520$ &  $5$ &  $3$ &  $-7190051054629560983266667941765809438720$\\
        $0$ &  $9$ &    $48611866333961358524429289555671840768$ &  $5$ &  $4$ &  $-5830328195321904270514091689296380559360$\\
        $0$ & $10$ &   $-23998790238070057635922084475102183168$ &  $5$ &  $5$ &  $-3069029408286947773330241492144734863360$\\
        $0$ & $11$ &     $8672731417323714657753514414490254592$ &  $5$ &  $6$ &  $-1190730817002543565227396363231795609600$\\
        $0$ & $12$ &    $-2075159091579767612621332730409169360$ &  $5$ &  $7$ &   $-320316632148217128204476180020466810880$\\
        $0$ & $13$ &      $322052239251448882191996011296477576$ &  $5$ &  $8$ &   $-103135907107074834693129446927731261440$\\
        $0$ & $14$ &      $-17556504634521010054008849438439143$ &  $5$ &  $9$ &    $-20078632448322012858575038616128389120$\\
        $1$ &  $0$ &    $-1938848215883310143996555439835709440$ &  $6$ &  $0$ &   $1101786715630392217099496609348856053760$\\
        $1$ &  $1$ &     $8208263455879762396473039394502082560$ &  $6$ &  $1$ &   $6644825759153207776628392157713929338880$\\
        $1$ &  $2$ &   $-58416152172890170054809898875743109120$ &  $6$ &  $2$ &   $9170724332114881662308034449179064401920$\\
        $1$ &  $3$ &    $75842480167323819896585446551600496640$ &  $6$ &  $3$ &  $10296333530288664715204547691370134896640$\\
        $1$ &  $4$ &  $-210775251847041676805670125392992665600$ &  $6$ &  $4$ &   $6843266110413098509844252043875922739200$\\
        $1$ &  $5$ &   $136184525798423937998405824948328202240$ &  $6$ &  $5$ &   $3030387074964732528675959937512201256960$\\
        $1$ &  $6$ &  $-196733800112955221741205021680955555840$ &  $6$ &  $6$ &    $915897437622215494357622042235575992320$\\
        $1$ &  $7$ &    $79216590016681069471177813962164797440$ &  $6$ &  $7$ &    $238947253619883730193810981276425912320$\\
        $1$ &  $8$ &   $-59859091103576645172144058190055751680$ &  $6$ &  $8$ &     $45054022158754454164122182527702794240$\\
        $1$ &  $9$ &    $18326220317126356156222467072302796800$ &  $7$ &  $0$ &  $-1854779415477316757095284416852388741120$\\
        $1$ & $10$ &    $-7108823201678288592667670159410595840$ &  $7$ &  $1$ &  $-7539087824448386866680226078048345128960$\\
        $1$ & $11$ &      $815712871940018411097123405288952320$ &  $7$ &  $2$ & $-11455283520474845151288748928631759175680$\\
        $1$ & $12$ &     $-361641384160782564047779894402070720$ &  $7$ &  $3$ & $-10532811484287412546421885016991177113600$\\
        $1$ & $13$ &      $-44637243978247300987553286056547600$ &  $7$ &  $4$ &  $-5628800651741725990749658960712486092800$\\
        $2$ &  $0$ &     $8645608055061412576173140279107256320$ &  $7$ &  $5$ &  $-1961746910015952161334849058844316794880$\\
        $2$ &  $1$ &     $8558501233501295140171817934064189440$ &  $7$ &  $6$ &   $-480465043427068303271618775842197340160$\\
        $2$ &  $2$ &   $208128644320609212070628856907148820480$ &  $7$ &  $7$ &    $-81866144453011763538579483060810547200$\\
        $2$ &  $3$ &    $25398307833670702694095833454451097600$ &  $8$ &  $0$ &   $2359500867194934432131610280534694952960$\\
        $2$ &  $4$ &   $607220768695981579445183976536172134400$ &  $8$ &  $1$ &   $7937559502154146780909418072268802621440$\\
        $2$ &  $5$ &    $19691331670165903708878704137721610240$ &  $8$ &  $2$ &  $10332903003270244092946854689083962163200$\\
        $2$ &  $6$ &   $391758194485028448586907909991164805120$ &  $8$ &  $3$ &   $7533871528692630975974582006997791539200$\\
        $2$ &  $7$ &     $6391741342689727853663049785223413760$ &  $8$ &  $4$ &   $3114332915955623314014735262982917324800$\\
        $2$ &  $8$ &    $51577861175147285247750616718405222400$ &  $8$ &  $5$ &    $807246277052955645212263471960754749440$\\
        $2$ &  $9$ &     $6204607039778934100147330129567334400$ &  $8$ &  $6$ &    $125533328851260714657310332124397568000$\\
        $2$ & $10$ &     $3640400071259879217670707810947143680$ &  $9$ &  $0$ &  $-2268800399225893262644750584090019758080$\\
        $2$ & $11$ &     $1983086792010676414706635015176821760$ &  $9$ &  $1$ &  $-6232822272707344104158292325400628428800$\\
        $2$ & $12$ &      $367557909233075714627665946035852480$ &  $9$ &  $2$ &  $-6605562463064358114347454886171377664000$\\
        $3$ &  $0$ &   $-39787816938051620059790474557930864640$ &  $9$ &  $3$ &  $-3636099546825763639762961786311855308800$\\
        $3$ &  $1$ &  $-159476071889802693109434895820400885760$ &  $9$ &  $4$ &  $-1071519765361097718706047123362467020800$\\
        $3$ &  $2$ &  $-731514012855923886206510772595248332800$ &  $9$ &  $5$ &   $-164378926637460663126315178916126392320$\\
        $3$ &  $3$ &  $-784400766293166137112748178029726924800$ & $10$ &  $0$ &   $1629123011829900148722393289644855787520$\\
        $3$ &  $4$ & $-1677552393625779813186347437480791244800$ & $10$ &  $1$ &   $3545142485298725935706734603329318420480$\\
        $3$ &  $5$ &  $-661353853184386013999740864276904017920$ & $10$ &  $2$ &   $2860895812933513581861598872551376814080$\\
        $3$ &  $6$ &  $-827343438367674818734162050362344734720$ & $10$ &  $3$ &   $1081718545854378956536728849551688990720$\\
        $3$ &  $7$ &   $-81305279726239321202490323786504601600$ & $10$ &  $4$ &    $178261545346595668666581910476742983680$\\
        $3$ &  $8$ &  $-105884795288052887319877149939803750400$ & $11$ &  $0$ &   $-849334966961448240072949565119885475840$\\
        $3$ &  $9$ &   $-12330592804548067461951671658298572800$ & $11$ &  $1$ &  $-1383440178129017513802206087174034554880$\\
        $3$ & $10$ &   $-11075426977582842938551752217607413760$ & $11$ &  $2$ &   $-760436696246979742345864896550356910080$\\
        $3$ & $11$ &    $-1828166883167359915600790033174784000$ & $11$ &  $3$ &   $-152903983934834948069172091852475596800$\\
        $4$ &  $0$ &   $160469345389784328942525434333096837120$ & $12$ &  $0$ &    $304611014439276620697260770439012024320$\\
        $4$ &  $1$ &   $716220525014818610326812472965385420800$ & $12$ &  $1$ &    $332812784800765996316435345702450626560$\\
        $4$ &  $2$ &  $2253008657830030943865379209044243251200$ & $12$ &  $2$ &     $94874071276729045364882989890351923200$\\
        $4$ &  $3$ &  $3005168691100945780240126435565921894400$ & $13$ &  $0$ &    $-67440944904415911980697770594884648960$\\
        $4$ &  $4$ &  $3634918639588564024870270462633338470400$ & $13$ &  $1$ &    $-37465022666738942217557941148492759040$\\
        $4$ &  $5$ &  $2039262448607305013342345116384399196160$ & $14$ &  $0$ &      $6976954565262648724575728467999457280$\\
        \hline
    \end{tabular}%
    }
\end{table*}

\section{Proofs of equivalences}
{%
\setlength{\arraycolsep}{.3em} 
\renewcommand\arraystretch{1}  

Two lattices with Gram matrices $\bA$ and $\bA'$ are equivalent
if and only if there exists a real positive scalar $c$
and an integer matrix $\bU$ with determinant $\pm 1$ such that
\begin{align}
\bA' = c \bU \bA \bU^\T .
\end{align}
To determine if two given lattices are equivalent is a nontrivial task.
In this appendix, we prove hitherto unknown equivalences
between known lattices, which were discovered during this work
and are mentioned in the text.
The search for constructive mapping rules between different lattice representations
was facilitated by a heuristic algorithm that selects a set of short vectors in each lattice
and attempts to map the sets onto each other.

\subsection{Equivalence between three $10$-dimensional lattices}
               
\label{app:equal10}
It was claimed in Sec.~\ref{sec:14Dident} that the top-left $10\times 10$ submatrix
of $\bB_{14}(\ba)$ generates a lattice that is equivalent to both
$K'_{10}$ in \cite[Sec.~8.5]{martinet03} and the lattice with automorphism group
$(C_6\times \mathrm{SU}_4(2))\colon C_2$ in \cite{souvignier94}.
Gram matrices of the two latter are available 
in the online Catalogue of Lattices \cite{nebe:catalog}
under the labels \emph{KAPPA10$'$} and \emph{(C6 x SU(4,2)):C2}, respectively, as
\begin{align}
\bA_1 &= \begin{bmatrix}
4 & -2 & -2 & 1 & 1 & 1 & -1 & -1 & -2 & -2 \\
-2 & 4 & 1 & -2 & -2 & -2 & -1 & -1 & 1 & 1 \\
-2 & 1 & 4 & -2 & -2 & -2 & 0 & 0 & 2 & 0 \\
1 & -2 & -2 & 4 & 1 & 1 & 0 & 0 & -1 & 0 \\
1 & -2 & -2 & 1 & 4 & 1 & 0 & 0 & -2 & -1 \\
1 & -2 & -2 & 1 & 1 & 4 & 2 & 2 & 0 & 0 \\
-1 & -1 & 0 & 0 & 0 & 2 & 4 & 1 & 0 & 0 \\
-1 & -1 & 0 & 0 & 0 & 2 & 1 & 4 & 2 & 2 \\
-2 & 1 & 2 & -1 & -2 & 0 & 0 & 2 & 4 & 1 \\
-2 & 1 & 0 & 0 & -1 & 0 & 0 & 2 & 1 & 4
\end{bmatrix}, \\
\bA_2 &= \begin{bmatrix}
4 & 1 & 1 & -2 & 1 & -2 & -2 & -2 & 2 & -2 \\
1 & 4 & 0 & -1 & 2 & 1 & 0 & 0 & 2 & -2 \\
1 & 0 & 4 & 1 & 2 & -2 & 1 & 1 & 2 & 1 \\
-2 & -1 & 1 & 4 & -1 & 1 & 2 & 1 & -1 & 2 \\
1 & 2 & 2 & -1 & 4 & -1 & 1 & 1 & 2 & 0 \\
-2 & 1 & -2 & 1 & -1 & 4 & 0 & 1 & -1 & 0 \\
-2 & 0 & 1 & 2 & 1 & 0 & 4 & 2 & -1 & 1 \\
-2 & 0 & 1 & 1 & 1 & 1 & 2 & 4 & -1 & 2 \\
2 & 2 & 2 & -1 & 2 & -1 & -1 & -1 & 4 & -1 \\
-2 & -2 & 1 & 2 & 0 & 0 & 1 & 2 & -1 & 4
\end{bmatrix}.
\end{align}
If $\bB_3$ denotes the top-left $10\times 10$ submatrix
of $\bB_{14}(\ba)$ in \eqref{eq:gen14}, then the equivalences can be explicitly shown as
\begin{align}
\bB_3 \bB_3^\T = \bU_1 \bA_1 \bU_1^\T = \bU_2 \bA_2 \bU_2^\T
\end{align}
with
\begin{align}
\bU_1 &= \begin{bmatrix}
1 & 0 & 0 & 0 & 0 & 0 & 0 & 0 & 0 & 0 \\
1 & 1 & 0 & 0 & 0 & 0 & 0 & 0 & 0 & 0 \\
1 & 1 & 1 & 1 & 0 & 0 & 1 & 0 & 0 & 0 \\
1 & 1 & 0 & 1 & 0 & -1 & 1 & 1 & 0 & 0 \\
-1 & -1 & 0 & 0 & -1 & 0 & -1 & 0 & -1 & 0 \\
-2 & -1 & -1 & 0 & -1 & 0 & -1 & 0 & -1 & -1 \\
1 & 0 & 1 & 0 & 0 & 0 & 0 & 0 & 0 & 0 \\
1 & 1 & 1 & 1 & 0 & 0 & 0 & 0 & 0 & 0 \\
0 & -1 & 0 & 0 & -1 & 0 & 0 & 0 & 0 & 0 \\
0 & -1 & -1 & 0 & -1 & -1 & 0 & 0 & 0 & 0
\end{bmatrix}, \\
\bU_2 &= \begin{bmatrix}
1 & 0 & 0 & 0 & 0 & 0 & 0 & 0 & 0 & 0 \\
0 & 0 & 0 & 0 & 0 & 0 & 0 & 0 & 1 & 0 \\
0 & -1 & 0 & 0 & 1 & 0 & 0 & 0 & 0 & 0 \\
1 & -1 & -1 & 0 & 0 & 0 & 1 & 1 & 1 & 0 \\
0 & -1 & -1 & 0 & 0 & 0 & 0 & 0 & 1 & 0 \\
0 & 0 & 1 & -2 & -1 & 1 & 1 & -1 & 0 & 1 \\
0 & 0 & -1 & 1 & 1 & -1 & -1 & 0 & 0 & 0 \\
0 & -1 & -1 & 0 & 1 & 0 & 0 & 0 & 1 & 0 \\
1 & -1 & -1 & 1 & 1 & 0 & 0 & 0 & 0 & 0 \\
1 & -1 & -1 & 0 & 0 & 0 & 1 & 0 & 1 & 0
\end{bmatrix}.
\end{align}

\subsection{Equivalence between three $12$-dimensional lattices}
\label{app:equal12}

It was claimed in Sec.~\ref{sec:intro} that the best $12$-dimensional lattices
in \cite{Kudryashov10} and \cite{agrell24K12} are equivalent.
Here we prove this equivalence, and also that the two best $12$-dimensional lattices
in \cite[Tab.~III]{Kudryashov10} are equivalent to each other.
Generator matrices for these two are
\begin{align}
\bB_4 &= \begin{bmatrix}
2 & 0 & 0 & 0 & 0 & 0 & 0 & 0 & 0 & 0 & 0 & 0 \\
0 & 2 & 0 & 0 & 0 & 0 & 0 & 0 & 0 & 0 & 0 & 0 \\
0 & 0 & 2 & 0 & 0 & 0 & 0 & 0 & 0 & 0 & 0 & 0 \\
0 & 0 & 0 & 2 & 0 & 0 & 0 & 0 & 0 & 0 & 0 & 0 \\
0 & 0 & 0 & 0 & 2 & 0 & 0 & 0 & 0 & 0 & 0 & 0 \\
0 & 0 & 0 & 0 & 0 & 2 & 0 & 0 & 0 & 0 & 0 & 0 \\
1 & 1 & 1 & 0 & 0 & 1 & 0 & 1 & 0 & 0 & 0 & 0 \\
0 & 0 & 1 & 1 & 1 & 0 & 0 & 1 & 0 & 1 & 0 & 0 \\
0 & 0 & 0 & 0 & 1 & 1 & 1 & 0 & 0 & 1 & 0 & 1 \\
0 & 1 & 0 & 0 & 0 & 0 & 1 & 1 & 1 & 0 & 0 & 1 \\
0 & 1 & 0 & 1 & 0 & 0 & 0 & 0 & 1 & 1 & 1 & 0 \\
1 & 0 & 0 & 1 & 0 & 1 & 0 & 0 & 0 & 0 & 1 & 1
\end{bmatrix}, \\
\bB_5 &= \begin{bmatrix}
2 & 0 & 0 & 0 & 0 & 0 & 0 & 0 & 0 & 0 & 0 & 0 \\
0 & 2 & 0 & 0 & 0 & 0 & 0 & 0 & 0 & 0 & 0 & 0 \\
0 & 0 & 2 & 0 & 0 & 0 & 0 & 0 & 0 & 0 & 0 & 0 \\
0 & 0 & 0 & 2 & 0 & 0 & 0 & 0 & 0 & 0 & 0 & 0 \\
0 & 0 & 0 & 0 & 2 & 0 & 0 & 0 & 0 & 0 & 0 & 0 \\
0 & 0 & 0 & 0 & 0 & 2 & 0 & 0 & 0 & 0 & 0 & 0 \\
1 & 1 & 1 & 1 & 1 & 1 & 1 & 0 & 1 & 1 & 0 & 0 \\
0 & 0 & 1 & 1 & 1 & 1 & 1 & 1 & 1 & 0 & 1 & 1 \\
1 & 1 & 0 & 0 & 1 & 1 & 1 & 1 & 1 & 1 & 1 & 0 \\
1 & 0 & 1 & 1 & 0 & 0 & 1 & 1 & 1 & 1 & 1 & 1 \\
1 & 1 & 1 & 0 & 1 & 1 & 0 & 0 & 1 & 1 & 1 & 1 \\
1 & 1 & 1 & 1 & 1 & 0 & 1 & 1 & 0 & 0 & 1 & 1
\end{bmatrix},
\end{align}
where the bottom half of each matrix contains a binary generator matrix for the
$[14,13]$ and $[37,35]$ tail-biting convolutional codes, respectively.
If $\bB_6$ denotes the generator matrix in \cite[Eq.~(13)]{agrell24K12}, then
 the equivalences can be explicitly shown as
\begin{align}
\bB_6 \bB_6^\T = \frac{1}{2} \bU_4 \bB_4 \bB_4^\T \bU_4^\T = \frac{1}{2} \bU_5 \bB_5 \bB_5^\T \bU_5^\T
\end{align}
with
\setlength{\arraycolsep}{.24em} 
\begin{align}
\bU_4 &= \begin{bmatrix}
2 & 1 & 1 & 0 & 0 & 1 & -2 & 0 & 0 & 0 & 0 & 0 \\
1 & 0 & 0 & 0 & 0 & 0 & 0 & 0 & 0 & 0 & 0 & 0 \\
2 & 1 & 1 & 0 & 0 & 2 & -3 & 1 & -1 & 1 & 0 & 0 \\
0 & 0 & 0 & 1 & 0 & -1 & 1 & -1 & 1 & -1 & 0 & 0 \\
2 & 2 & 1 & -1 & -1 & 2 & -4 & 2 & -1 & 1 & -1 & 1 \\
2 & 2 & 1 & -1 & 0 & 2 & -4 & 2 & -1 & 1 & -1 & 1 \\
2 & 2 & 0 & -1 & -1 & 2 & -4 & 3 & -1 & 1 & -1 & 1 \\
0 & 0 & -1 & 0 & 0 & 0 & 0 & 0 & 0 & 0 & 0 & 0 \\
-1 & -1 & -1 & 0 & 0 & -2 & 2 & -1 & 1 & -1 & 1 & 0 \\
-1 & -1 & -1 & 0 & 0 & -1 & 2 & -1 & 1 & -1 & 1 & 0 \\
-1 & -1 & -1 & 1 & 0 & -2 & 3 & -2 & 2 & -1 & 1 & -1 \\
-2 & -3 & -1 & 2 & 1 & -4 & 6 & -5 & 3 & -2 & 2 & -1
\end{bmatrix}, \\
\bU_5 &= \begin{bmatrix}
1 & 0 & 0 & 0 & 0 & -1 & 0 & 0 & 0 & 0 & 0 & 0 \\
1 & 0 & 0 & 0 & 0 & 0 & 0 & 0 & 0 & 0 & 0 & 0 \\
0 & 0 & 0 & 0 & 0 & 0 & 0 & -1 & 0 & 1 & 0 & 0 \\
1 & 0 & 0 & 0 & -1 & -1 & 0 & 1 & 0 & -1 & 0 & 0 \\
1 & 1 & 0 & 0 & 0 & -1 & 0 & 1 & 0 & 0 & 0 & -1 \\
0 & 0 & 0 & 0 & 0 & 0 & 0 & -1 & 0 & 0 & 0 & 1 \\
0 & 0 & -1 & -1 & -1 & 0 & 1 & 0 & -1 & 0 & 0 & 1 \\
-1 & -2 & -1 & -1 & -2 & -1 & 2 & 0 & 0 & -2 & 0 & 2 \\
-1 & -2 & -1 & -1 & -2 & -1 & 1 & 0 & 0 & -2 & 1 & 2 \\
-1 & -2 & -1 & 0 & -2 & -1 & 1 & 0 & 0 & -2 & 1 & 2 \\
0 & 1 & -1 & -1 & 1 & 1 & 0 & 0 & -1 & 2 & -1 & 0 \\
0 & -1 & 0 & 0 & -1 & -1 & 0 & 0 & 1 & -1 & 0 & 1
\end{bmatrix}.
\end{align}
}

\end{document}